\documentclass[a4paper,fleqn,usenatbib]{mnras}

\usepackage[T1]{fontenc}
\usepackage{ae,aecompl}

\usepackage{amsmath}
\usepackage{amssymb}
\usepackage{natbib}
\usepackage{graphicx}
\usepackage{xcolor}

\title[A MUSE study of RCS 0224]{A gravitationally-boosted MUSE survey for emission-line galaxies at $z\gtrsim5$ behind the massive cluster RCS\,0224\thanks{Partially based on observations made with the NASA/ESA Hubble Space Telescope, obtained at the Space Telescope Science Institute, which is operated by the Association of Universities for Research in Astronomy, Inc., under NASA contract NAS 5-26555. These observations are associated with program \#14497.} }
\author[R. Smit et al.]{Renske Smit,$^{1}$ 
A.M. Swinbank,$^{1,2}$ 
Richard Massey,$^{1,2}$ 
Johan Richard,$^{3}$ 
\newauthor Ian Smail,$^{1,2}$ 
 and J.-P. Kneib$^{4,5}$ \\
$^{1}$Centre for Extragalactic Astronomy, Durham University, South Road, Durham DH1 3LE UK\\
$^{2}$Institute for Computational Cosmology, Durham University, South Road, Durham DH1 3LE UK\\
$^{3}$Univ Lyon, Univ Lyon1, Ens de Lyon, CNRS, Centre de Recherche Astrophysique de Lyon UMR5574, F-69230, Saint-Genis-Laval, France\\
$^{4}$Institute of Physics, Laboratory of Astrophysics, Ecole Polytechnique F\'{e}d\'{e}rale de Lausanne (EPFL), Observatoire de Sauverny, 1290 Versoix, Switzerland\\
$^{5}$Aix Marseille Universit\'{e}, CNRS, LAM (Laboratoire d'Astrophysique de Marseille) UMR 7326, 13388, Marseille, France\\
}
\begin{document}

\date{Accepted 2017 January 26. Received 2017 January 16; in original form 2016 September 4}

\pubyear{2017}


\maketitle

\begin{abstract}

We present a VLT/MUSE survey of lensed high-redshift galaxies behind the $z=0.77$ cluster RCS\,0224$-$0002. 
We study the detailed internal properties of a highly magnified  ($\mu\sim29$)
$z = 4.88$ galaxy seen through the cluster. We detect wide-spread nebular \ion{C}{iv}$\lambda\lambda$1548,1551{\,\AA} emission from this galaxy as well as a bright Ly$\alpha$ halo with a spatially-uniform wind and absorption profile across 12 kpc in the image plane. 
Blueshifted high- and low-ionisation interstellar absorption indicate the presence of a high-velocity outflow ($\Delta v\sim300\,\rm km\,s^{-1}$) from the galaxy. 
Unlike similar observations of galaxies at $z\sim2-3$, the Ly$\alpha$ emission from the halo emerges close to the systemic velocity - an order of magnitude lower in velocity offset than predicted in ``shell''-like outflow models. To explain these observations we favour a model of an outflow with a strong velocity gradient, which changes the effective column density seen by the Ly$\alpha$ photons.  
We also search for high-redshift Ly$\alpha$ emitters and identify 14 candidates between $z=4.8-6.6$, including an over-density at $z=4.88$, of which only one has a detected counterpart in  $HST$/ACS+WFC3 imaging.  

\end{abstract}


\section{Introduction}

Over the last decade,  deep observations of blank fields, in particular with the \textit{Hubble Space Telescope (HST)} have identified a substantial poplation of galaxies beyond $z>3$, using broadband photometry \citep[e.g.][]{Madau1996,Steidel1996,Steidel1999,Sawicki1997,Lehnert2003,Giavalisco2004,Ouchi2004,Mclure2009,vanderBurg2010,Bowler2015,Bouwens2015,Finkelstein2015}. 
Despite the progress in identifying large numbers of galaxies, it remains challenging to obtain spectroscopic redshifts and determine the physical properties of these systems. This is largely due to their inherent faintness and the fact that bright rest-frame optical emission-line tracers such as H$\alpha$ and [\ion{O}{iii}], which are traditionally used to measure the properties of the ISM, are shifted to observed mid-infrared wavelengths for sources at $z\gtrsim3-4$.  The small physical sizes of galaxies at $z>3$ compared to typical ground-based seeing also makes spatially resolved observations difficult to obtain, inhibiting measurements of dynamical masses, star-formation distributions and wind energetics.

Recently, the commissioning of the Multi Unit Spectroscopic Explorer (MUSE) on the Very Large Telescope (VLT) has led to an advance in the identification and characterisation of $z\sim3-6$ galaxies though wide-field and deep spectroscopy of the rest-frame ultraviolet (UV) spectra of these sources. 
For example, MUSE is starting to probe the physical properties of \ion{H}{ii} regions within galaxies by exploiting gravitationally lensing through their faint UV nebular emission lines such as \ion{C}{iv}$\lambda\lambda$1548,1551{\,\AA}, \ion{He}{ii}$\lambda$1640{\,\AA}, \ion{O}{iii}]$\lambda\lambda$1661,1666{\,\AA} and \ion{C}{iii}]$\lambda\lambda$1907,1909{\,\AA} \citep[][]{Karman2015,Caminha2016,Patricio2016,Vanzella2016}, lines which are rarely seen in local star-forming galaxies \citep[e.g.][]{Hainline2011,Rigby2015}. These lines are produced either by young, metal-poor stellar populations with high-ionization parameters \citep[e.g.][]{Christensen2012,Stark2014,Rigby2015}, or gas photo-ionisation by faint active galactic nuclei \citep[AGN; e.g.][]{Stark2015,Feltre2016}. Furthermore, MUSE has enabled the detailed modelling of extended Ly$\alpha$ emission, gaining insights into the inflowing neutral gas and/or wind energetics in the circum-galactic medium (CGM) of galaxies \citep{Swinbank2015,Bina2016,Gullberg2016,Wisotzki2016,Vanzella2017}.  

Moreover, MUSE is a promising new instrument for undertaking unbiased spectroscopic surveys. \citet{Bacon2015} used a 27 hour MUSE pointing of the $Hubble$ Deep Fields South (HDF-S)
to detect 89 Lyman-$\alpha$ emitters in the redshift range $z\sim3-6$.
Remarkably, 66\% of the Ly$\alpha$ emitters above $z\gtrsim5$ have
no counterpart in the \emph{HST} broadband imaging (to a
limiting magnitude of $m_i\sim29.5$), 

In this paper, we extend current work on characterising the UV spectra of intrinsically faint high-redshift galaxies out to $z\sim5$ through the analysis of VLT/MUSE observations of one of the most strongly magnified galaxies known above a redshift of $z>3$; the highly magnified ($\mu=13-145\times$) $z=4.88$  lensed arc seen through the core of the compact $z=0.77$ cluster  RCS\,0224$-$0002 \citep[hereafter S07]{Gladders2002,Swinbank2007}.

S07 observed nebular [\ion{O}{ii}] emission and an extended Ly$\alpha$ halo  in this $z=4.88$ source and they hypothesized that a galactic-scale bipolar outflow has recently bursted out of this system and into the intergalactic medium (IGM). Our new observations obtain significantly higher signal-to-noise ratio (S/N) in the UV emission and continuum, allowing us to resolve the shape of the Ly$\alpha$ profile and detect the UV-interstellar medium (ISM)
lines.  Furthermore, our MUSE pointing covers the complete $z\sim6$ critical curves, which allows for an efficient survey for faint high-redshift Ly$\alpha$ emitters. These sources are important targets to study in order to understand the properties of the ultra-faint galaxy population that could have contributed significantly to reionisation. 

This paper is organised as follows: we describe our MUSE dataset and we summarize the complementary data presented by S07 in \S\ref{sec:data}. We analyse the spectral properties of the main $z=4.88$ arc in \S\ref{sec:results}. We present the results of a blind search for Ly$\alpha$ emitters in \S\ref{sec:blind_search} and finally we summarise our findings in \S\ref{sec:summary}.  

For ease of comparison with previous studies we take $H_0=70\,\rm km\,s^{-1}\,Mpc^{-1},\,\Omega_{\rm{m}}=0.3,\,$and$\,\Omega_\Lambda=0.7$, resulting in an angular scale of 6.4 kpc per arcsecond at $z=4.88$. Magnitudes are quoted in the AB system \citep{OkeGun}.

\begin{figure*}
\includegraphics[scale=0.85,trim=3mm 105mm 0mm 95mm]{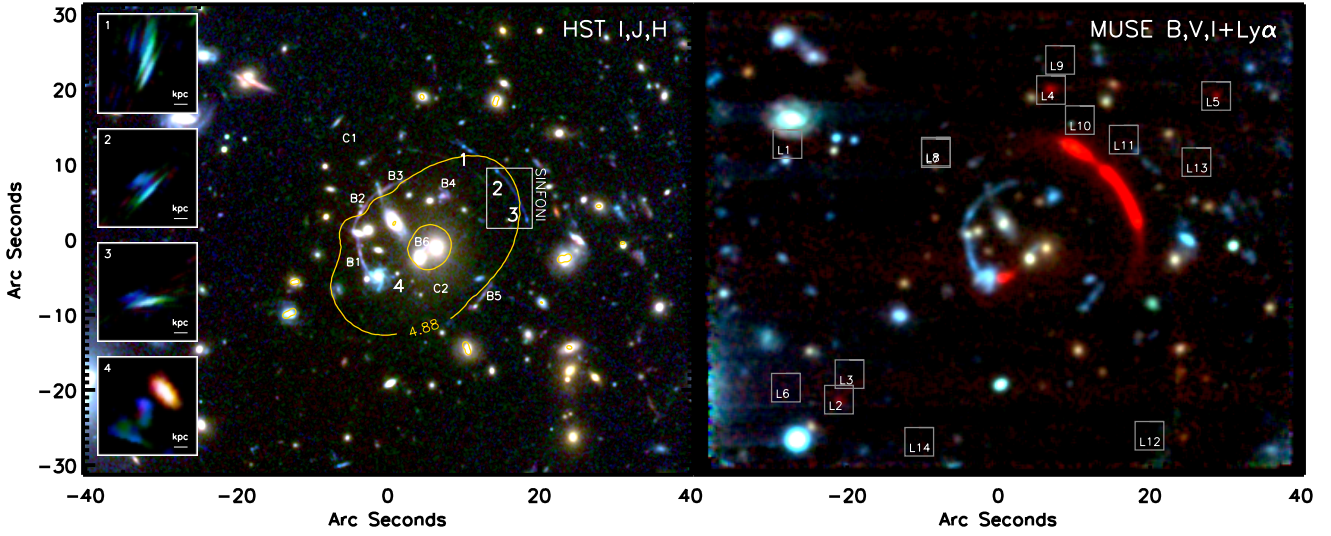}
\caption{ $HST$ $I_{814}$, $J_{125}$ and $H_{160}$ (left panel)  and  MUSE $B,V,$\,$I$+7143{\,\AA} (right panel) colour images of the cluster core of RCS\,0224$-$0002 at $z$\,=\,0.77. The four images of the lensed  galaxy at $z=4.88$ are numbered 1--4 in the left panel and are reconstructed in the source plane in the inset panels. We also indicate the coverage of the SINFONI K-band spectroscopy  and the images B1-B6 of a $z_{\ion{C}{iii}]}=2.4$ galaxy and images C1-C2 of a Ly$\alpha$ emitter at $z_{\rm Ly\alpha}=5.5$ that are used as constraints on the lens model (\S\ref{sec:lensmodel}). The $z=4.8-6.5$ Ly$\alpha$ candidates selected in \S\ref{sec:blind_search} are marked with squares on the right panel (see \S \ref{sec:blind_search} and Appendix \ref{app:lyaem}). For the MUSE red channel we combine a broad-band centered on 6125{\,\AA} with an 8{\,\AA}-wide narrowband centered on the Ly$\alpha$ halo around the $z\sim4.88$ arc \citep{Gladders2002}; the Ly$\alpha$ emission is clearly extended beyond the continuum. Three sources (L2, L4 and L5) in our blind search for Ly$\alpha$ emitters (\S\ref{sec:blind_search}) are found at the same redshift as the $z=4.88$ arc, suggesting a galaxy over-density. 
   }
\label{fig:RCS0224_color}
\end{figure*}

\begin{figure*}
\includegraphics[scale=.9,trim=25mm 150mm 20mm 40mm]{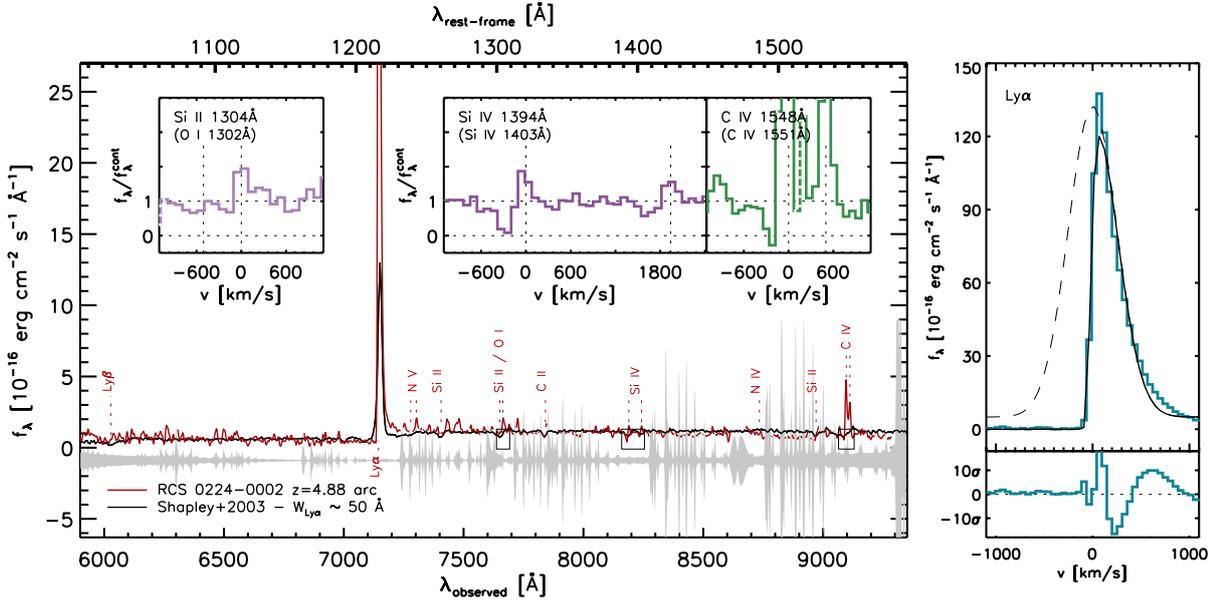}
\caption{\textit{Left panel:} The  MUSE one-dimensional spectrum extracted over the stellar continuum of the $z=4.88$ arc (red line), smoothed with a Gaussian filter with $\sigma=3${\,\AA} (the grey filled region shows the sky spectrum, offset from the spectrum for clarity). Regions with strong sky emission lines are omitted for clarity. The solid black line indicates the stacked spectrum of the strongest Ly$\alpha$ emitting Lyman break galaxies in the \citet{Shapley2003} sample. The spectrum is normalised to match the $z=4.88$ arc continuum at $\sim$7500 {\,\AA} and redshifted to $z_{[\ion{O}{ii}]}=4.8757$. The inset panels  show the spectra with respect to the [\ion{O}{ii}] redshift for the \ion{Si}{ii}$\lambda$1304{\,\AA}, \ion{Si}{iv}$\lambda$1394{\,\AA} and \ion{C}{iv}$\lambda$1548{\,\AA} lines (indicated with black squares on the main spectrum). The spectra are binned to two times lower spectral sampling (2.5{\,\AA}/pix) and the position of the \ion{O}{i}$\lambda$1302{\,\AA}, \ion{Si}{ii}$\lambda$1304{\,\AA}, \ion{Si}{iv}$\lambda\lambda$1394,1403{\,\AA}  and \ion{C}{iv}$\lambda\lambda$1548,1551{\,\AA} lines are indicated with dashed lines. The parts of the spectra strongly affected by skylines are indicated with a dashed line in the insets. We detect a single-peaked, strongly asymmetric Ly$\alpha$  line,  narrow and strong \ion{C}{iv} emission and narrow blueshifted \ion{C}{iv} and \ion{Si}{iv} absorption lines. All absorption features are signficantly blueshifted with respect to the [\ion{O}{ii}] redshift. \textit{Right panel:} Zoom-in of the observed Ly$\alpha$ line (blue line, top panel) and a Gaussian with Voigt-profile absorber fit to the data (black lines). The narrow Ly$\alpha$ peak in combination with the high-velocity tail are not well described by this simple model as shown by the residuals (bottom panel). 
   }
\label{fig:RCS0224_spec}
\end{figure*}

\section{Data}
\label{sec:data}

\subsection{\textit{HST} imaging}
We obtained $HST$ imaging from the Space Telescope Science Institute  MAST data archive (GO:14497, PI: Smit and GO: 9135, PI:Gladdders). RCS\,0224$-$0002 ($\alpha = 02$:24:34.26, $\delta= -$00:02:32.4) was observed with the Wide Field Planetary Camera 2 (WFPC2) using the F666W ($V_{666}$) filter (10.5 ks), the Advanced Camera for Surveys (ACS) using the F814W ($I_{814}$) filter (2.2 ks) and the Wide Field Camera 3 (WFC3) using the F125W ($J_{125}$) and F160W ($H_{160}$) filters (2.6 ks each). The ACS and WFC3 images were reduced with Drizzlepac v2.1.3 to 0.05 and 0.128 arcsec pixel$^{-1}$ resolution respectively. The depth of the $I_{814}$, $J_{125}$ and $H_{160}$ band images is 26.3, 26.8 and 26.7 mag respectively ($5\sigma$ in a 0.5\arcsec-diameter aperture). The WFPC2 data was reduced with the {\small STSDAS} package from {\small IRAF} to $\sim0.1$ arcsec pixel$^{-1}$ resolution as described by S07. A false-color image using the $I_{814}$, $J_{125}$ and $H_{160}$ bands are shown in Figure \ref{fig:RCS0224_color}. The color image shows two bright arcs at  $z=2.40$ (lensed images B1--B6) and $z=4.88$ (lensed images 1--4).  

\subsection{MUSE spectroscopy}

We observed the cluster RCS\,0224$-$0002  with a single pointing ($\sim1\times 1$ arcmin) of the VLT/MUSE IFU spectrograph \citep{Bacon2010} between November 13, 2014, and September 16, 2015, programme 094.A-0141. Each individual exposure was 1500 seconds, with spatial dithers of $\sim$15 arcsec to account for cosmic rays and defects. One observing block was partly taken in twilight and therefore omitted from the final data-cube, resulting in a co-added exposure time of 13.5 ks. All the observations we use were taken in dark time with $<$0.8" $V$-band seeing and clear atmospheric conditions. 

We reduced the data with the public MUSE {\small ESOREX} pipeline version 1.2.1, including bias, dark, flat-fielding, sky subtraction, wavelength and flux calibrations. For each individual exposure we used the lamp flat-field taken adjacent in time to the observation for illumination correction. The reduced data-cubes were registered and stacked using the {\small EXP\_COMBINE} routine.
 The seeing measured on the combined exposure is $\sim$0.68" full width half max (FWHM), with a spectral resolution of 94 $\rm km\,s^{-1}$ (2.2 \,\AA) FWHM at 7000\,\AA. 
 
 A false-color image constructed from the final MUSE cube is shown in Figure \ref{fig:RCS0224_color}. We use median images centered on 5375{\,\AA}, 6125{\,\AA} and 8275{\,\AA} as broadband inputs and we add a 8{\,\AA} wide mean image centered on 7146{\,\AA} to the red channel to emphasize the Ly$\alpha$ emission in the $z=4.88$ arc. All bright $HST$ sources are detected in the MUSE continuum, while the $z=4.88$ arc is clearly detected with spatially extended Ly$\alpha$ emission. A number of other Ly$\alpha$ sources are identified at the same redshift (see \S\ref{sec:blind_search}).

\subsection{SINFONI spectroscopy}
To complement the MUSE dataset we exploit the SINFONI IFU spectroscopy presented by S07. Briefly, the SINFONI data was taken in the $HK$ grating ($\lambda/\Delta\lambda = 1700$) covering the [\ion{O}{ii}]$\lambda\lambda$3726.1,3728.8{\,\AA} doublet redshifted to $\sim2.2\,\mu$m. The $\sim8\times 8$ arcsec field-of-view (with a spatial resolution 0.25 arcsec pixel$^{-1}$) covers the lensed images 2  and 3 of the $z=4.88$ arc .

\section{Analysis and Discussion}
\label{sec:results}

\subsection{Lens model}
\label{sec:lensmodel}
To constrain the intrinsic properties of the emission-line galaxies in this study we require an accurate lens-model. S07 constructed a simple mass-model of RCS\,0224$-$0002 with the two main elliptical galaxies in the centre of the cluster and the dark matter component approximated by single truncated pseudo-isothermal elliptical mass distributions. Their primary observational constraints on the mass configuration are the four lensed images of the $z=4.88$ arc. However, our MUSE observations also cover the other arcs in the cluster. We extract spectra over the multiply-imaged central blue arcs (B1--B6 in Figure \ref{fig:RCS0224_color}) and detect \ion{C}{iii}]$\lambda\lambda$1907,1909{\,\AA}, \ion{O}{iii}]$\lambda\lambda$1661,1666{\,\AA} emission and \ion{Si}{ii}$\lambda$1403{\,\AA}, \ion{Si}{iv}$\lambda\lambda$1394,1403{\,\AA}, \ion{Si}{ii}$\lambda$1526{\,\AA}, \ion{C}{iv}$\lambda\lambda$1548,1551{\,\AA}, \ion{Fe}{ii}$\lambda1608${\,\AA}, \ion{Al}{ii}$\lambda$1671{\,\AA} absorption in images B1--B6 and measure a redshift $z_{\ion{C}{iii}]}=2.396\pm0.001$ from the integrated light of these images (Smit et al., in preparation). 

We use these new constraints to update the lens model by S07.  As in S07, the lens modelling is performed using the {\small LENSTOOL} software \citep{Kneib1996,Jullo2007,Jullo2009}. {\small LENSTOOL} is a parametric  method for  modelling  galaxy clusters that uses a Markov Chain Monte Carlo (MCMC) fit for a specified number of mass peaks. Each mass peak corresponds to a dark matter halo  modelled with a truncated pseudo-isothermal elliptical that is characterised by a position ($\rm RA, dec$), velocity dispersion $\sigma_{\rm V}$, ellipticity $\varepsilon$, truncation radius $r_{\rm cut}$ and core radius $r_{\rm core}$.

For our updated mass model we include mass components for the brightest 22 cluster members and two components for the cluster halo.
We include constraints from the six images B1--B6 of the $z=2.396$ galaxy arc, including the de-magnified image in the centre. Another faint arc is identified in the $HST$ imaging (D1--D3 in Table~\ref{tab:lens_constraints}), just inside the $z=4.88$ arc, but we do not detect any emission lines from this source in the MUSE data-cube. Furthermore, we search the MUSE cube for bright multiply lensed line emitters and find a Ly$\alpha$ emitter without an $HST$ continuum counterpart at $z_{\rm Ly\alpha}=5.500 \pm 0.002$  (labelled C1 at $\alpha = 02$:24:34.86, $\delta= -00$:02:16.2 and C2 at $\alpha = 02$:24:34.02, $\delta= -00$:02:36.3 in Figure \ref{fig:RCS0224_color}). The locations of the Ly$\alpha$ emitter images are well predicted by the lens-model that uses all other constraints and therefore we include this doubly lensed image as an additional constraint. In Figure \ref{fig:RCS0224_color} we show the critical curve of our new model at a redshift of $z=4.88$ and we list all multiple images used to constrain the model in Table~\ref{tab:lens_constraints} in Appendix~\ref{app:table_lensmodel}. 

Our mass model differs from that of S07 in three ways. First, owing to the different assumed cosmology, our model is $\sim35$\% less massive: $M=(3.8\pm0.2)\times10^{14}$M$_\odot$ compared to $M=(5.9\pm0.4)\times10^{14}$M$_\odot$. We recover the S07 mass if we switch back to their cosmology. Second, the inclusion of mass from all cluster member galaxies makes the $z=2.4$ critical line better match the observed features of lensed system B. Third, our distribution of mass is more elongated toward the North-West. The S07 model had close to circular symmetry, forced by a prior on the ellipticity of the cluster-scale dark matter. This resulted in a scatter of rms$^A_i=1.21\arcsec$ between the predicted and observed positions of images A1-A4 (G.\ Smith et al.\ private comm.). By dropping the prior (and simultaneously imposing constraints from newly identified lens systems), a cluster-scale mass distribution with ellipticity $\varepsilon=0.63$ achieves rms$_i^A=0.52\arcsec$, or rms$_i=1.03\arcsec$ for all image systems. However, we achieve a still-better fit (rms$_i^A=0.48\arcsec$, rms$_i=0.88\arcsec$) using two cluster-scale halos. These were given a Gaussian prior centred on the two BCGs. The first gets asymmetrically offset to the North-West; the second remains near CG2. This two-halo model achieves a superior log(Likelihood) of --26.57 and $\chi^2=61.3$ in 11 degrees of freedom compared to the best-fitting one-halo model, which has log(Likelihood) of --142.92 and $\chi^2=294$ in 17 degrees of freedom. The best-fit parameters are listed in Table~\ref{tab:lensparam} in Appendix~\ref{app:table_lensmodel}.

  \begin{table}
\centering
\caption{Detected spectral features of the $z=4.88$ arc (integrated over galaxy images 1, 2 and 3)}
\begin{tabular}{lccc}
\hline
\hline

line & $z$ & $\Delta v$~[km$\,\rm s^{-1}$]$\,^a$ & EW$_0$~[\AA] \\
\hline
  \multicolumn{4}{c}{Emission lines}\\ 
\hline
Ly$\alpha^{\rm halo}$ & 4.8770$\pm$0.0005$\,^b$ & 68$\pm$37 & 278$\pm$55 \\
Ly$\alpha^{\rm cont}$ & 4.8770$\pm$0.0005$\,^b$ & 68$\pm$37 & 135$\pm$27 \\
\ion{Si}{ii}$ \lambda$1304{\,\AA}  & 4.8761$\pm$0.0006  &  21$\pm$41  &  1.1$\pm$0.1 \\ 
\ion{Si}{iv}$\lambda$1394{\,\AA}  & 4.8738$\pm$0.0003 & -96$\pm$29 & 0.6$\pm$0.1 \\
\ion{Si}{iv}$\lambda$1403{\,\AA}  & 4.8752$\pm$0.0006 & -27$\pm$39 &  0.6$\pm$0.1 \\
\ion{C}{iv}$\lambda$1548{\,\AA}  & 4.8753$\pm$0.0001  & -23$\pm$26 & 5.6$\pm$0.4 \\
\ion{C}{iv}$\lambda$1551{\,\AA}  & 4.8755$\pm$0.0001  & -12$\pm$26  & 3.7$\pm$0.3\\
\hline
  \multicolumn{4}{c}{Absorption lines}\\ 
\hline
\ion{Si}{ii}$ \lambda$1304{\,\AA}$\,^c$  & 4.8710$\pm$0.0001  & -240$\pm$25  &  -0.2$\pm$0.1 \\ 
\ion{Si}{iv}$\lambda$1394{\,\AA}  & 4.8694$\pm$0.0001  & -322$\pm$26 &  -1.1$\pm$0.1  \\
\ion{Si}{iv}$\lambda$1403{\,\AA}  & 4.8693$\pm$0.0002  & -327$\pm$28 &  -0.5$\pm$0.1  \\
\ion{C}{iv}$\lambda$1548{\,\AA}  & 4.8694$\pm$0.0001  &  -322$\pm$26 &  -0.8$\pm$0.1 \\
\hline

\end{tabular}
\flushleft
$^a$ Velocity offset with respect to the the systemic redshift $z_{[\ion{O}{ii}]}=4.8757\pm0.0005$. Uncertainties combine the uncertainty on the line redshift with the uncertainty on the [\ion{O}{ii}] redshift. 
$^b$ Using the peak of the Ly$\alpha$ line. 
$^c$ Blended with the \ion{O}{i}$\lambda$1302{\,\AA} line. 

\label{tab:z4.88lines}
\end{table}

\subsection{The $z=4.88$ arc}
\label{sec:arc}

The $z=4.88$ arc was first discovered  in the Red-Sequence Cluster (RCS) by \citet{Gladders2002}. \citet{Gladders2002} detected the bright Ly$\alpha$ emission in galaxy images 1--3 at $z_{\rm Ly\alpha}=4.8786$ with VLT/FORS-2 spectroscopy. S07 targeted the arc with VLT/VIMOS (galaxy images 1--4) and VLT/SINFONI  (galaxy images 2--3) spectroscopy and detected Ly$\alpha$ at $z_{\rm Ly\alpha}=4.8760$ and [\ion{O}{ii}]$\lambda\lambda$3726.1,3728.8{\,\AA} at $z_{ [\ion{O}{ii}]}=4.8757$. S07 measured  a star-formation rate of $12\pm2\rm\,M_\odot\, yr^{-1}$, a velocity gradient of $\lesssim60\rm\,km\,s^{-1}$, and an estimated dynamical mass of $\sim 10^{10}\rm\, M_\odot$ within 2\,kpc from the [\ion{O}{ii}] emission lines.

For our MUSE study of the $z=4.88$ arc we will assume the systemic velocity of the galaxy is best estimated by $z_{\rm sys}=z_{ [\ion{O}{ii}]}=4.8757\pm0.0005$ (integrated over galaxy images 2--3). Furthermore, from our lens-model we find luminosity weighted amplifications of $\mu=29^{+9}_{-11}$, $\mu=21^{+12}_{-8}$, $\mu=138^{+7}_{-74}$ and $\mu=1.30^{+0.01}_{-0.01}$ for images 1, 2, 3 and 4 respectively (note that image 3 has a very high amplification, but also a very large uncertainty, because the arc crosses the critical curve).  These values are slightly higher than the mean, luminosity-weighted magnification of $\mu=16\pm2$ found by S07 for images 1,2 and 3 integrated (though within the uncertainties for images 1 and 2). The uncertainty on our numbers is largely due to the fact that a small shift of the critical curve can change the luminosity weighted amplification significantly. In particular, we note that the high magnification of image 3 is dominated by a few pixels that overlap with the critical curve, while the estimated magnification for any modelling method is most uncertain near the critical curves \citep[see][]{Meneghetti2016}.     

To measure the detailed properties of the UV spectrum of this galaxy we first construct a one-dimensional spectrum (up to $\sim1600${\,\AA} in the rest-frame) of the $z=4.88$ arc from the MUSE cube by measuring the integrated (non-weighted) spectrum extracted from pixels in the lensed images 1, 2 and 3 with a S/N>2$\sigma$ in the continuum image of the MUSE data-cube. The resulting  spectrum is shown in Figure \ref{fig:RCS0224_spec}. As well as bright Ly$\alpha$ emission, which has an observed equivalent width (EW) of $793\pm159${\,\AA} (rest-frame EW$_0=135\pm27${\,\AA}), we clearly detect the absorption line doublet \ion{Si}{iv}$\lambda\lambda$1394,1403{\,\AA}, which originates in the ISM and/or CGM and the emission line doublet \ion{C}{iv}$\lambda\lambda$1548,1551{\,\AA}, with some evidence for an absorption component as well (see inset panels), which is likely to arise from a combination of stellar, nebular and ISM/CGM components. The observational parameters of the UV spectroscopic features in the MUSE data are listed in Table \ref{tab:z4.88lines} (see Appendix \ref{app:table_images} for measurements on the individual lensed galaxy images).

In the next sections we will first discuss the morphology of the emission lines, before moving to a detailed analysis of the spectral properties of the $z=4.88$ arc, the kinematics of the system and the physical picture that emerges from these observations.

\begin{figure*}
\includegraphics[scale=1.3,trim=60mm 192mm 60mm 35mm]{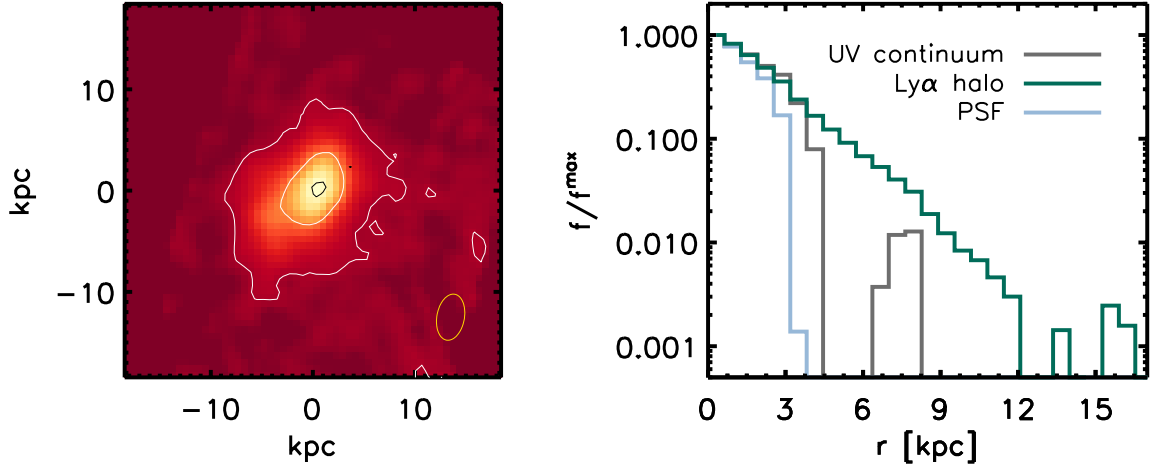}
\caption{\textit{Left:} The source-plane reconstruction of the Ly$\alpha$ halo of the lensed galaxy image 4. White contours indicate 50\% and 10\% of the peak flux in the Ly$\alpha$ halo, while the yellow ellipse in the corner indicates the 50\% peak flux for the de-lensed PSF of the MUSE continuum image. The black contour indicates 50\% of the peak flux of the UV continuum observed in the $HST$ $I_{814}$ band. The Ly$\alpha$ flux is dominated by a smooth and largely circularly symmetric halo that is extended beyond the stellar continuum, but we also identify an extended region to the South-East (North-West in the image plane).
 \textit{Right:} The exponential surface-brightness profile of the Ly$\alpha$ halo, compared to the MUSE continuum (using a fore-ground subtracted broad-band image at 7400{\AA}) surface-brightness profile and the MUSE PSF measured from the source-plane reconstruction of  galaxy image 4. The Ly$\alpha$ halo is extended beyond the stellar component measured from the UV continuum.
   }
\label{fig:RCS0224_sourceplane}
\end{figure*}

\begin{figure}
\center
\includegraphics[scale=1.,trim=30mm 168mm 95mm 35mm]{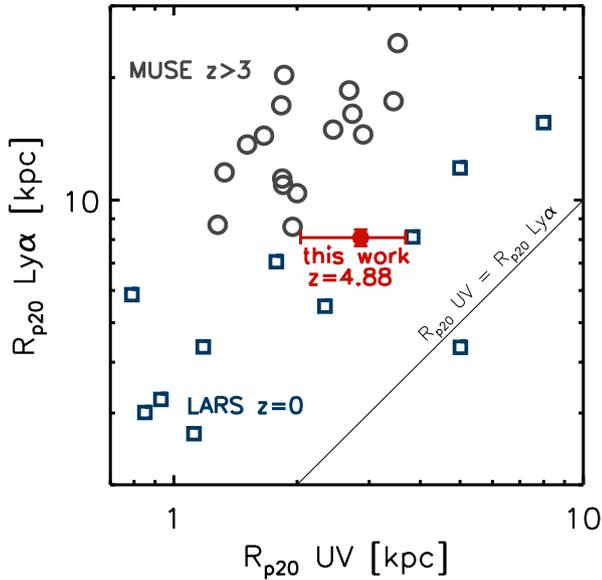}
\caption{A comparison of the Ly$\alpha$ and UV continuum Petrosian scale radii for the $z=4.88$ arc and including measurements at $z=0-6$ from the Ly$\alpha$ reference sample  \citep[LARS;][]{Hayes2013} and MUSE HDF-S observations \citep{Wisotzki2016}.
}
\label{fig:RCS0224_petrosian}
\end{figure}

\begin{figure}
\includegraphics[scale=1.35,trim=82mm 180mm 0mm 50mm]{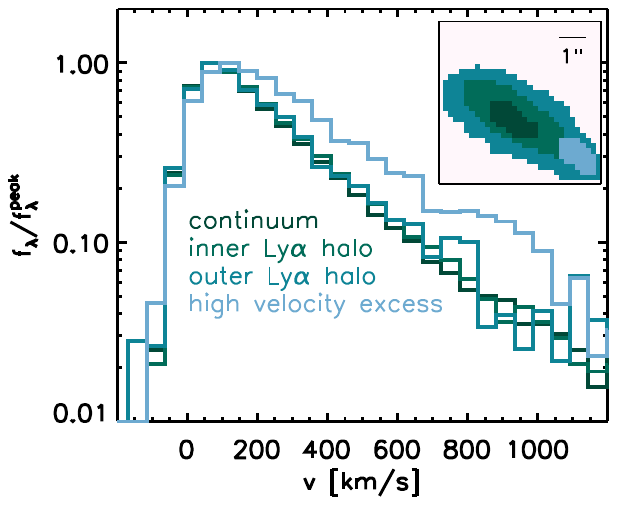}
\caption{The spatial dependence of the Ly$\alpha$ velocity profile (with respect to the [\ion{O}{ii}] redshift) in different bins in the image plane of galaxy image 1. The black line indicates the flux within a contour over the stellar continuum flux and the green lines are emitted within contours of constant Ly$\alpha$ flux. The light blue line indicates a region of high-velocity excess flux. The pixels used to generate the four spectra are shown in the inset panel. The spatial variation of the Ly$\alpha$ halo is surprisingly small, with the peak of the Ly$\alpha$ line varying less than the width of one spectral element corresponding to $\rm\sim60\,km\,s^{-1}$. The small difference in the velocity profile between the inner and outer halo, would suggest that the bulk of the Ly$\alpha$ photons are CGM generated/rescattered, since there is no evidence for an ISM component that falls away as we observe Ly$\alpha$ further away from the stellar continuum. Spatially offset from the spherical halo, we detect excess flux out to 1000\,km s$^{-1}$ (corresponding to the extended region to the South-East in the source-plane image shown in Figure \ref{fig:RCS0224_sourceplane}), possibly indicating a collimated high velocity outflow superimposed on an isotropic component that dominates the total Ly$\alpha$ emission (see \S\ref{sec:lya}). 
   }
\label{fig:RCS0224_speclya}
\end{figure}

\subsubsection{Ly$\alpha$ morphology}
\label{sec:lya}

The Ly$\alpha$ emission in the $z=4.88$ arc (see Figure \ref{fig:RCS0224_color}) appears to be significantly extended.
Lyman Break galaxies and Ly$\alpha$ emitters at $z\sim2-6$ often exhibit extended Ly$\alpha$ halos around the stellar continuum of the galaxies, following an exponential surface brightness distribution \citep{Steidel2011,Matsuda2012,Momose2014,Wisotzki2016}. These  Ly$\alpha$ halos are thought to be generated either by cooling radiation \citep[e.g.,][]{Dijkstra2006,Dijkstra2009,Faucher2010,Rosdahl2012} or by resonant scattering from a central powering source, such as star-formation or AGN \citep[e.g.,][]{Verhamme2006,Gronke2015}.
First, we  investigate the morphology of Ly$\alpha$ in the $z=4.88$ arc behind RCS\,0224$-$0002. 

Figure \ref{fig:RCS0224_sourceplane} shows the source-plane reconstruction of the continuum subtracted Ly$\alpha$ halo. We use image 4, since the Ly$\alpha$ halos of images 1 to 3 are incomplete and merged together (see the right panel of Figure \ref{fig:RCS0224_color}).  For the spatial profile we use bins of 0.1 arcsec in concentric circles around the peak flux of the Ly$\alpha$ emission.  The MUSE Ly$\alpha$ halo has an observed FWHM of 2.2 kpc, while the $HST$ continuum has a FWHM of 0.2 kpc. A number of foreground cluster galaxies contaminate the measurement of the stellar spatial profile directly from the continuum image. Therefore we construct a broadband image redwards of Ly$\alpha$, centered on 7400{\AA}, and we subtract a continuum image bluewards of the Ly$\alpha$ break centered on  6975{\AA} in order to remove most of the foreground contamination. We mask any remaining flux from foreground sources by hand. Furthermore, we extract the PSF from a nearby star and place this at the position of the Ly$\alpha$ peak in our lens model in order to construct the source-plane image of the PSF and measure its spatial profile.  

The Ly$\alpha$ halo appears roughly isotropic, with little substructure, except for an extended lower luminosity region in the South-East. Comparing the Ly$\alpha$ halo with the UV continuum image in Figure \ref{fig:RCS0224_sourceplane} indicates the extended nature of the faint Ly$\alpha$ profile beyond the continuum. The Ly$\alpha$ halo is consistent with an exponential profile. For comparison with Ly$\alpha$ halos in the literature, we measure the Petrosian radius \citep{Petrosian1976} of the halo, defined as the annulus where the Ly$\alpha$ flux is equal to $\eta$ times the mean flux within the annulus. The Petrosian radius is a useful measure, since it is only weakly dependent on the seeing of the observations. For $\eta=20\%$ we find $R_{\rm p20,Ly\alpha}=8.1\pm0.4\,\rm kpc$ and $R_{\rm p20,UV}=2.9\pm0.8\,\rm kpc$, which is somewhat lower than the range of Petrosian radii $R_{\rm p20,Ly\alpha}\sim10-30\,\rm kpc$ (for $R_{\rm p20,UV}\sim1.3-3.5\,\rm kpc$) found by \citet{Wisotzki2016}, but similar to some of the largest Ly$\alpha$ halos around $z\sim0$ analogues in the LARS sample \citep{Hayes2013}, which typically show radii $R_{\rm p20,Ly\alpha}\lesssim8\,\rm kpc$ (Figure \ref{fig:RCS0224_petrosian}). 

\citet{Faucher2010} and \citet{Rosdahl2012} use radiative transfer simulations to investigate the expected morphology of Ly$\alpha$ halos generated by cooling radiation and they predict concentrated emission, which can extend out to $\rm 10-30 \rm\,kpc$.  While stacked Ly$\alpha$ halos extend out to $\sim 100\, \rm kpc$ \citep{Steidel2011,Matsuda2012,Momose2014}, indicating cooling is not the origin of Ly$\alpha$ emission in typical galaxies, our observations do not have sufficient S/N to trace the $z=4.88$ Ly$\alpha$ halo beyond $\rm 10\, kpc$, necessary to rule out a gas cooling scenario. We will therefore further investigate whether the faint, extended Ly$\alpha$ emission is produced by resonant scattering from a central source or by cooling radiation in \S\ref{sec:spec_lya} based on the spectral properties of the line.

\begin{figure}
\includegraphics[scale=1.,trim=35mm 180mm 100mm 37mm]{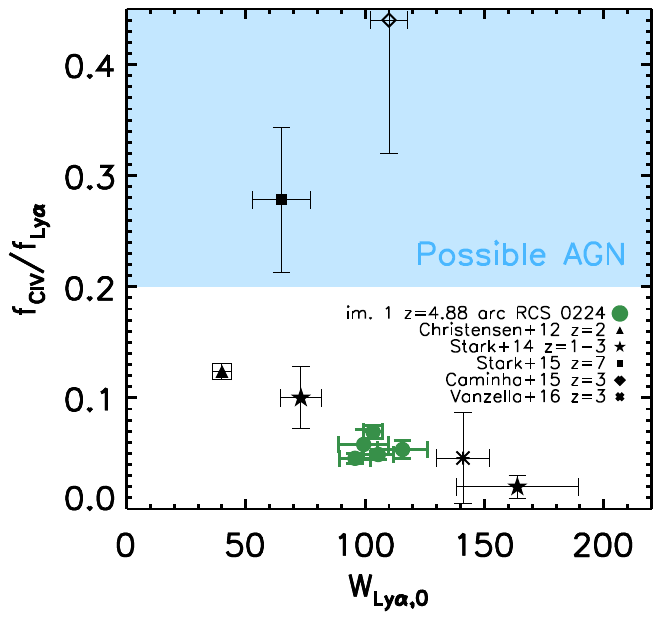}
\caption{ Comparison of the emission line ratio  \ion{C}{iv}/Ly$\alpha$ as a function of rest-frame Ly$\alpha$ equivalent width with similar detections in the literature. We show the measurements along the lensed image 1 of the $z=4.88$ arc. We also show data from lensed  \ion{C}{iv} emitters at $z\sim2$ found by \citet{Christensen2012}, \citet{Stark2014}, \citet{Stark2015}, \citet{Caminha2016} and \citet{Vanzella2016}. Narrow-line AGN have been observed to show a emission line ratio of \ion{C}{iv}/Ly$\alpha\gtrsim0.2$ \citep{Shapley2003,Erb2010,Hainline2011}. Along the $z=4.88$ arc this emission line ratio is steady and always below \ion{C}{iv}/Ly$\alpha<0.1$ (see \S\ref{sec:CIV}). 
   }
\label{fig:RCS0224_CIVLya}
\end{figure} 
 
\begin{figure*}
\includegraphics[scale=0.75,trim=0mm 5mm 20mm 150mm]{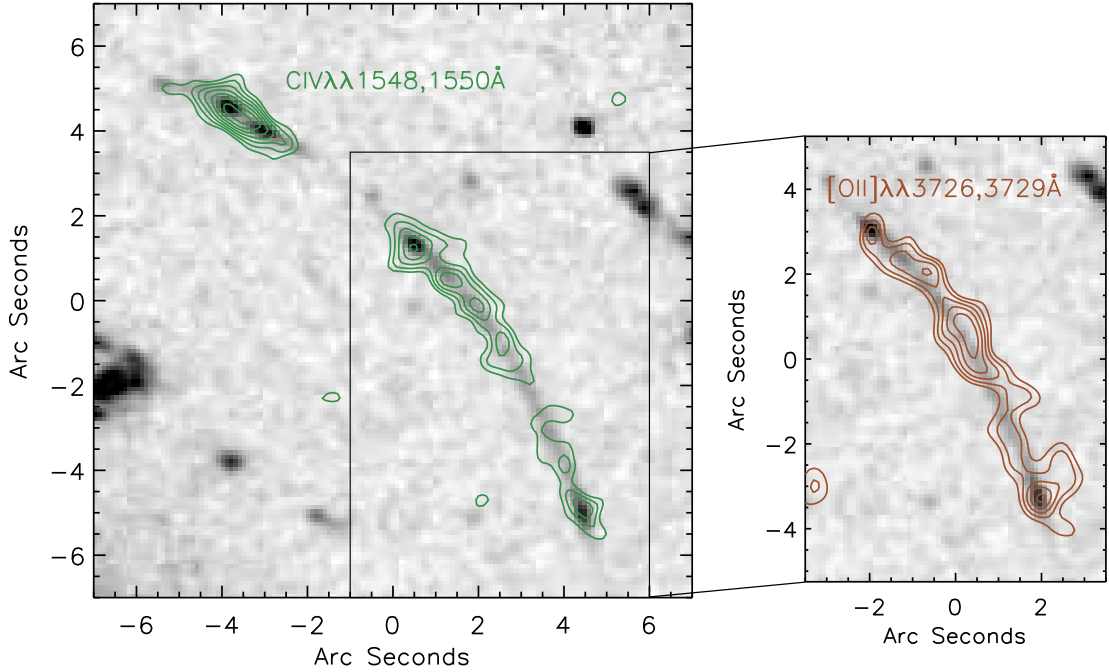}
\caption{The spatial distribution of the \ion{C}{iv} emission lines along the lensed galaxy images 1, 2 and 3 of the $z=4.88$ arc (green contours, left panel), compared to the [\ion{O}{ii}] distribution over the lensed images 2 and 3 from the SINFONI data (S07, brown contours, right panel). The $HST$ ($V_{666}+I_{814}$)-band image is shown in greyscale. The high-ionisation emission lines are spatially extended along the arc and trace the UV-continuum light. The morphologies of \ion{C}{iv} and [\ion{O}{ii}] show similarities that might imply a nebular origin of \ion{C}{iv} in a large number of the star-forming regions of the $z=4.88$ arc.  
   }
\label{fig:RCS0224_CIVcontour}
\end{figure*}

\begin{figure*}
\includegraphics[scale=0.99,trim=70mm 180mm 60mm 35mm]{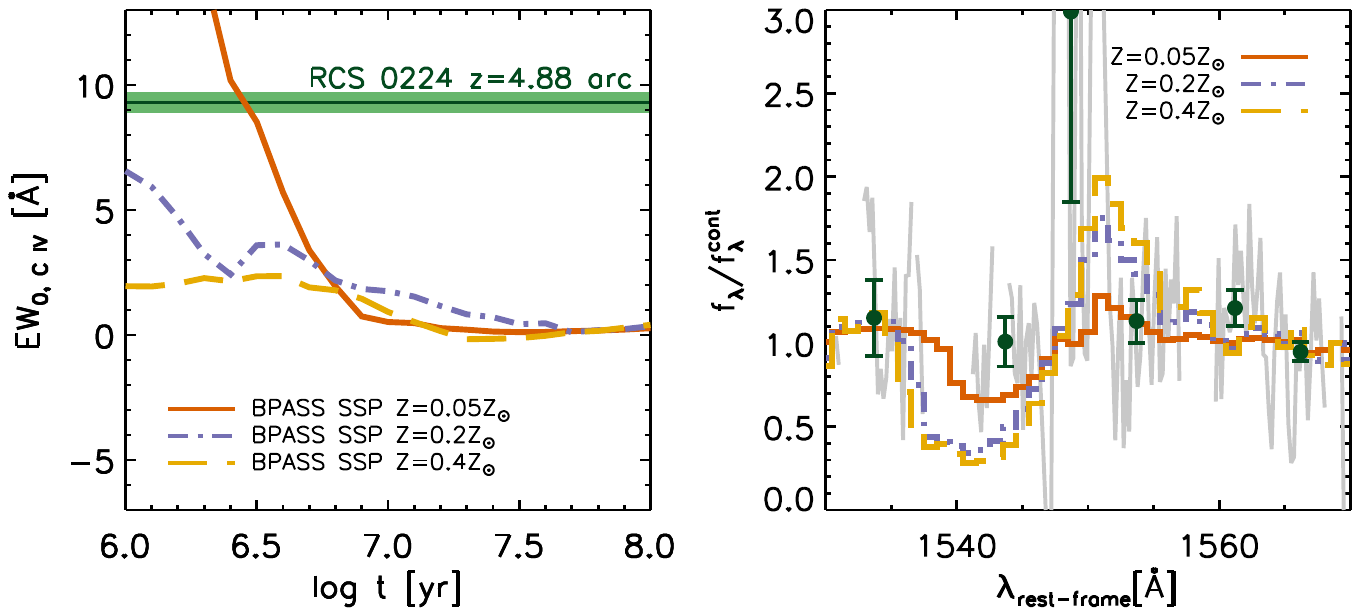}
\caption{  \textit{Left:} Predicted \ion{C}{iv} equivalent width evolution of nebular \ion{C}{iv} emission during the first 100 Myr of a single stellar population at $Z=0.05,0.2,0.4\,Z_\odot$ metallicity (solid, dot-dashed and dashed lines respectively) using the BPASS stellar population synthesis models including binary rotation \citep{Eldridge2012}. The green region indicates the rest-frame equivalent width observed (including uncertainty) in the integrated spectrum of the $z=4.88$ arc. The nebular \ion{C}{iv} emission could be produced by a very low-metallicity stellar population if the star clusters  are observed during the first $\sim3$ Myr of their lifetime. \textit{Right:} A comparison of the stellar \ion{C}{iv} profiles (excluding nebular emission) predicted by the same three BPASS models at 3 Myr with our data. The binned points on either side of the nebular \ion{C}{iv} lines show no evidence for the strong P-Cygni profiles produced by stellar winds, indicating a metallicity of the stellar population lower than the $Z=0.05\,Z_\odot$ BPASS model. 
   }
\label{fig:RCS0224_CIVEW}
\end{figure*} 

\subsubsection{Spectral properties of the Ly$\alpha$ line}
\label{sec:spec_lya}

High-redshift  Ly$\alpha$ emitters can exhibit a wide range of spectral properties, such as blueshifted and redshifted emission, single and double peaked lines and different line widths and velocity offsets, which gives insight into the emission mechanism of Ly$\alpha$ and the column density and velocity distribution of the ISM and CGM neutral gas \citep[e.g.][]{Verhamme2006,Gronke2015}.

 For the $z=4.88$ arc, the Ly$\alpha$ emission line profile is very asymmetric and we find a single redshifted Ly$\alpha$ line, with a peak at $z_{\rm Ly\alpha}=4.8770\pm0.0005$ (using the wavelength and width of the spectral element where Ly$\alpha$ peaks), $\sim40-90\rm\,km\,s^{-1}$ redshifted with respect to the [\ion{O}{ii}] emission which marks the systemic redshift, and $\rm FWHM_{red}=285\,km\,s^{-1}$, with very little flux bluewards of the [\ion{O}{ii}] redshift (see Figures \ref{fig:RCS0224_spec} and \ref{fig:RCS0224_speclya}). We set an upper limit on the presence of a weaker blue line; at $-v_{\rm red}$ we find an upper limit on the flux ratio of any blue peak to the red peak $F_{\rm peak,blue}/F_{\rm peak,red}<0.027$. Furthermore, we detect a faint tail of redshifted Ly$\alpha$ emission out to $\sim1000\rm\,km\,s^{-1}$. A simplified model for this asymmetric line shape is that of a Gaussian emission line profile convolved with a Voigt profile, describing the collisional and Doppler broadening of interstellar absorption lines, as shown in Figure \ref{fig:RCS0224_spec}, where we fix the redshift of the underlying Gaussian emission to the [\ion{O}{ii}] redshift $z=4.8757$. The best-fit model in Figure \ref{fig:RCS0224_spec} indicates a \ion{H}{i} absorber with a column density of $10^{19}\,\rm cm^{-2}$, however, the fit fails to reproduce both the narrow peak and the high-velocity tail of the Ly$\alpha$ line. In fact, the emission-line component of the Ly$\alpha$ line shows a strongly non-gaussian shape; instead we observe an exponential profile as a function of velocity over two orders of magnitude in flux (Figure \ref{fig:RCS0224_speclya}), remarkably similar to the exponential surface brightness profile (Figure \ref{fig:RCS0224_sourceplane}).

S07 observed the modest redshifted  narrow Ly$\alpha$ line in combination with the high velocity tail, and interpreted this as a combination of emission from the central source combined with redshifted emission from an outflow. To test this model, in Figure \ref{fig:RCS0224_speclya} we show the spatial variation of the spectral Ly$\alpha$ profile in the image-plane of the lensed galaxy image 1 (the highest S/N image). While we used the source-plane reconstruction of galaxy image 4  for deriving the spatial properties of the Ly$\alpha$ halo, we use the brightest galaxy image for spectral analysis to obtain higher signal-to-noise information. First, we partition the halo along contours of constant observed Ly$\alpha$ flux. While the Ly$\alpha$ flux in the halo drops by more than an order of magnitude compared to the emission over the stellar continuum, the shape of the Ly$\alpha$ profile, after normalising to the peak flux, changes only marginally from the Ly$\alpha$ profile extracted over the stellar continuum. Across the lensed image, the wavelength of the peak of the Ly$\alpha$ line changes by less than  $\sim50\rm\,km\,s^{-1}$, while the width and the shape of the high velocity tail stays nearly constant \citep[see also][for a similar pattern in a Ly$\alpha$ halo at $z=3.51$]{Patricio2016}. 

These results differ strongly from the scenario described in S07, where the main peak of the Ly$\alpha$ profile comes directly from the star-forming regions and the high velocity wing is re-scattered in an expanding shell of gas within the CGM. For this model to hold we would expect the star-formation component (the peak of Ly$\alpha$) to drop off rapidly with increasing radius, while the back-scattered CGM component changes little with radius, and therefore the peak flux would shift to higher velocities and the shape would change significantly. The spatially-uniform Ly$\alpha$ spectral profile instead suggests that the Ly$\alpha$ peak is also produced or resonantly scattered within the CGM, which generates  Ly$\alpha$ emission with a wide range in velocities.

We can test this further by searching for any deviations from the average Ly$\alpha$ profile. We  use the spectrum extracted over the stellar continuum as a model for fitting the Ly$\alpha$ line in each individual pixel of the $z=4.88$ arc, leaving the normalisation as the only free parameter and considering only the peak of the Ly$\alpha$ line as a model constraint. After subtracting our one-parameter model we detect only a weak residual. In  Figure \ref{fig:RCS0224_speclya} we show the spectrum extracted over the region with the largest residual, which shows a slightly offset peak compared to the Ly$\alpha$ extracted over the stellar continuum and a broadened profile out to $1000\rm\,km\,s^{-1}$, indicating a collimated high-velocity outflow on top of the isotropic Ly$\alpha$ halo component that is described above. 

Given the extended nature of the Ly$\alpha$ emission (\S\ref{sec:lya}) we now consider various generation mechanisms for the emission. Given the spatially invariant Ly$\alpha$ line profile, which indicates that only a minor fraction of the Ly$\alpha$ emission reaches us directly from the galaxy, it is reasonable to consider whether the halo can be produced by cooling radiation from the CGM. \citet{Dijkstra2006} and \citet{Faucher2010} model such scenarios using radiative transfer simulations and find that Ly$\alpha$ should typically be double peaked and blueshifted with respect to the systemic velocity of the galaxy. Assuming these models provide a reasonable description of the system, the single redshifted Ly$\alpha$ peak we observe excludes cooling as a source of Ly$\alpha$ photons in the $z=4.88$ arc. 

To reproduce the Ly$\alpha$ line profile for the $z=4.88$ arc we thus favour a picture where a central powering source is surrounded by a largely isotropic halo of neutral gas, which dampens Ly$\alpha$ bluewards of the systemic velocity and resonantly scatters the vast majority of photons towards higher velocities within the expanding gas behind the galaxy \citep[e.g.][]{Dijkstra2006,Verhamme2006}. 

The strong similarity between the Ly$\alpha$ surface brightness profile (Fig. \ref{fig:RCS0224_sourceplane}) and the spectral profile in the $z=4.88$ arc could suggest the presence of a smoothly varying velocity gradient in the CGM gas that resonantly scatters the Ly$\alpha$ photons into our line of sight. This scenario is qualitatively in good agreement with the Ly$\alpha$ profile considered in \citet[][see their Fig. 7]{Verhamme2006}, shown in Appendix \ref{app:lyaprof}, where the relatively low column density (at a given velocity) created by the strong velocity gradient causes  the escape of Ly$\alpha$ photons predominantly at low velocity (and small radii), while a weak high-velocity tail is still observed due to the photons that are resonantly scattered through the accelerating outflow. A model with a gas velocity gradient furthermore predicts the absence of a blue peak, which is difficult to reproduce in a shell model, since a low velocity shell with low covering fraction which gives rise to a red Ly$\alpha$ peak close to the systemic velocity also produces a nearly symmetric blue peak. 
In \S\ref{sec:discussion} we will discuss how this model fits into a physical picture that can explain our full set of observations.

\subsubsection{C\,{\scriptsize IV} emission}
\label{sec:CIV}
The detection of narrow ($\rm FWHM = 156\pm 16\rm\, km\,s^{-1}$) \ion{C}{iv}$\lambda\lambda$1548,1551{\,\AA} emission in the $z=4.88$ arc is interesting since the UV spectra of field galaxies generally show  \ion{C}{iv} in \textit{absorption} from ISM/CGM gas, or else exhibit a P-Cygni profile from the stellar winds of O-stars, with \ion{C}{iv} emission redshifted by a few hundred $\rm km\,s^{-1}$ \citep{Leitherer2001,Shapley2003,Jones2012}.  AGN can also produce  \ion{C}{iv} in emission, though with typical line-widths of at least a few hundred $\rm km\,s^{-1}$. Narrow \ion{C}{iv}$\lambda\lambda$1548,1551{\,\AA} has so far been observed in a handful of strongly lensed high-redshift galaxies \citep[e.g.][]{Holden2001,Christensen2012,Stark2014,Stark2015,Caminha2016,Vanzella2016}. To date these galaxies have either been studied with slit spectroscopy or they are unresolved  in ground-based observations, inhibiting the study of the spatial distribution of the \ion{C}{iv}$\lambda\lambda$1548,1551{\,\AA}. The MUSE observations of the brightly lensed $z=4.88$ arc therefore provides us with a unique opportunity to investigate the origin of this line in more detail.

In the absence of rest-frame optical spectroscopy, a common approach to assessing the possible presence of AGN is using UV emission line ratios \citep[e.g.,][]{Stark2015,Feltre2016}. At $z=4.88$, this requires near-infrared spectroscopy to measure the \ion{He}{ii}$\lambda$1640{\,\AA} and \ion{C}{iii}]$\lambda\lambda$1907,1909{\,\AA} lines. With the current observations we can only assess the \ion{C}{iv}/Ly$\alpha$ ratio, which has a ratio of $\gtrsim0.2$ in the composite spectra of AGN \citep{Shapley2003,Hainline2011}. In contrast, this ratio is \ion{C}{iv}/Ly$\alpha =0.054\pm0.006$ in the $z=4.88$ arc, with little variation along the images, consistent with the interpretation that this line is associated with star formation and not with a hidden AGN. 
Figure \ref{fig:RCS0224_CIVLya} shows the \ion{C}{iv}/Ly$\alpha$ ratio as a function of Ly$\alpha$ equivalent width for various  lensed galaxies in the literature \citep{Christensen2012,Stark2014,Stark2015,Caminha2016,Vanzella2016} and in 5 spatial bins along the lensed image 1 of the $z=4.88$ arc (using both the 1548 and 1551{\,\AA} lines).  The \ion{C}{iv}/Ly$\alpha$ ratio and the observed equivalent width of \ion{C}{iv} $55\pm2${\,\AA} (rest-frame EW$_0=9.3\pm0.4${\,\AA}) do not change significantly as a function of position along galaxy image 1. Furthermore, we observe no strong emission from typical AGN lines such as \ion{N}{v}$\lambda$1240{\,\AA}, \ion{S}{iv}$\lambda\lambda$1393,1402{\,\AA} and  \ion{N}{iv}$\lambda\lambda$1483,1486{\,\AA} \citep[e.g.,][]{Hainline2011}.

In Figure \ref{fig:RCS0224_CIVcontour} we show the spatial distribution of \ion{C}{iv}$\lambda\lambda$1548,1551{\,\AA}, using a continuum subtracted narrowband image of the \ion{C}{iv} emission and overlaying the contours on the $HST$ continuum image.  \ion{C}{iv} clearly extends along the arc and shows a morphology that is consistent with the [\ion{O}{ii}] emission. While we would expect a centrally concentrated source for \ion{C}{iv} if it was originating from an AGN, we can distinguish at least four different `clumps' in the \ion{C}{iv} morphology with similar brightness,  suggesting that the  \ion{C}{iv} emission is nebular in origin and emerging from multiple star-forming regions throughout the galaxy. 

Finally, we measure the UV-continuum slope $\beta$ ($f_\lambda \propto \lambda^\beta$)  from the $J_{125}-H_{160}$ color of galaxy image 1 (integrated flux) and find  $\beta=$-$2.19\pm0.14$, while the individual star-forming clumps along the arc show slopes between $\beta$=-$1.68-$-2.64. Since high-redshift galaxies that host faint AGN have measured UV continuum slopes of $\beta\sim$-1.4$-$-0.3 \citep{Hainline2011,Giallongo2015} and we therefore conclude that low-mass accreting black holes are unlikely to contribute to the radiation field giving rise to the \ion{C}{iv} emission.

\subsubsection{Metal absorption lines}
\label{sec:spec_metals}

The high-ionisation \ion{C}{iv}$\lambda\lambda$1548,1551{\,\AA} line is expected to be a combination of nebular, stellar and ISM/CGM components.
While no absorption by the ISM appears present at the systemic velocity of  \ion{C}{iv}, blueshifted \ion{C}{iv} absorption is observed at $\Delta v=-322\pm26\rm\, km\,s^{-1}$ which also has a narrow profile, with $\rm FWHM\sim200\, km\,s^{-1}$. 

The spectrum displays a strong similarity between the \ion{C}{iv}$\lambda$1548{\,\AA} and \ion{Si}{iv}$\lambda$1394{\,\AA} absorption profiles  (Figure \ref{fig:RCS0224_spec}), indicating the absorption of both lines is due to highly ionised gas clouds in the ISM/CGM of the galaxy moving towards us. 
Furthermore, both  \ion{Si}{iv}$\lambda$1394{\,\AA} and \ion{Si}{ii}$\lambda$1304{\,\AA} show no evidence for absorption at the systemic velocity (Figure \ref{fig:RCS0224_spec}) and we even find weak emission lines, possibly indicating a low covering fraction of gas in the ISM of the galaxy. Given the strong absoprtion of the high-ionisation lines at $\sim-300\rm\, km\,s^{-1}$, galactic feedback in this galaxy has possibly ejected a large fraction of the interstellar gas into the CGM/IGM.

The \ion{Si}{ii}$\lambda$1304{\,\AA} absorption appears to be weaker than  the \ion{C}{iv}$\lambda$1548{\,\AA} and \ion{Si}{iv}$\lambda$1394{\,\AA} absorption lines. The ratio of the equivalent with of the \ion{Si}{ii}$\lambda$1304{\,\AA} line over that of \ion{Si}{iv}$\lambda$1394{\,\AA} is EW(\ion{Si}{ii}$\lambda$1304{\,\AA})/EW(\ion{Si}{iv}$\lambda$1394{\,\AA})=0.2 (see Table \ref{tab:z4.88lines}). 
This is in contrast to the typical UV spectra of high-redshift galaxies, where the low-ionisation lines are stronger than the high-ionisation lines of the same species, with for example EW(\ion{Si}{ii}$\lambda$1304{\,\AA})/EW(\ion{Si}{iv}$\lambda$1394{\,\AA})=1.2 in the composite spectrum of Lyman break galaxies by \citet{Shapley2003}.
We also fit the low- and high-ionisation lines with a Gaussian profile convolved with a Voigt-profile absorber and estimate column densities of $\log{N/\rm cm^{-2}}=14.5\pm0.3$ and $\log{N/\rm cm^{-2}}=14.6\pm0.1$ respectively.
 Possibly a larger fraction of the outflowing gas is highly ionised due to a hard ionisation field, which is present given the widespread nebular  \ion{C}{iv}$\lambda\lambda$1548,1551{\,\AA} emission in the galaxy. 
It is possible that the neutral gas swept up by the ionised outflow is optically thin because of this, or else that the covering fraction of neutral gas in the CGM is incomplete \citep[e.g.,][]{Erb2015}. To distinguish between these explanations we would need a clean observation of the \ion{Si}{ii}$\lambda$1260{\,\AA} and \ion{Si}{ii}$\lambda$1527{\,\AA} absorption features, which are currently obscured by skylines.

\begin{figure}
\includegraphics[scale=1.0,trim=30mm 145mm 50mm 35mm]{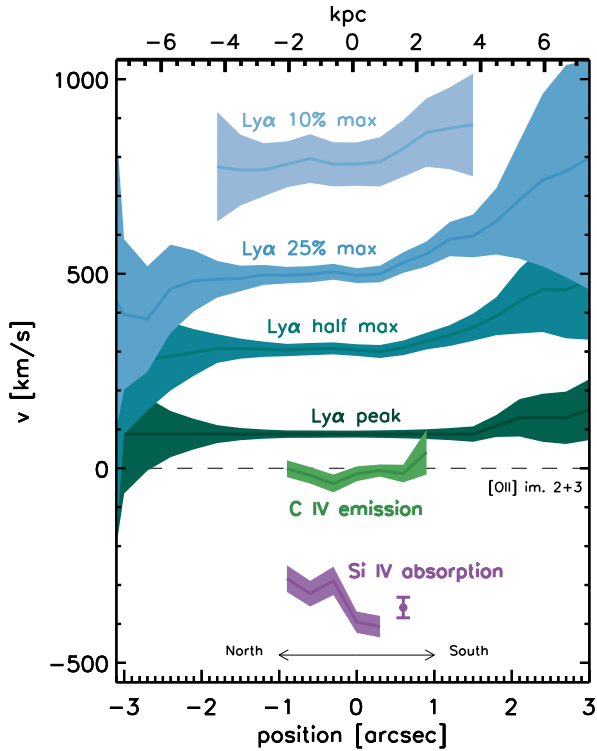}
\caption{The spatial variation of the various line components along a tangential line  through lensed image 1 of the $z=4.88$ arc (from North-East to South-West in the image plane, corresponding to a line from North to South in the source plane) with respect to the [\ion{O}{ii}] redshift of $z=4.8757$ measured from galaxy images 2 and 3. The \ion{C}{iv} emission is close to the [\ion{O}{ii}] redshift, while the \ion{Si}{iv} interstellar absorption lines are blue-shifted by $\sim300-400\rm \,km\,s^{-1}$ (the solid purple point corresponds to a measurement combining galaxy images 1, 2 and 3 to obtain a $>5\sigma$ line detection). The peak of the Ly$\alpha$ halo is redshifted by $\sim80\rm \,km\,s^{-1}$, with very little velocity structure along the arc. The width the Ly$\alpha$ line increases on one side of the galaxy as indicated by the positions of the half, 25\%, and 10\% of the max flux in the line. We find no evidence of a strong velocity gradient in the nebular lines, and the peak of the Ly$\alpha$ emission stays nearly constant even beyond the galaxy. There is, however, a widening of the Ly$\alpha$ profile and a weak velocity gradient in the \ion{Si}{iv} absorption, possibly indicating a collimated high-velocity outflow. 
   }
\label{fig:RCS0224_velpos}
\end{figure}

\begin{figure}
\center
\includegraphics[scale=1.,trim=30mm 168mm 95mm 35mm]{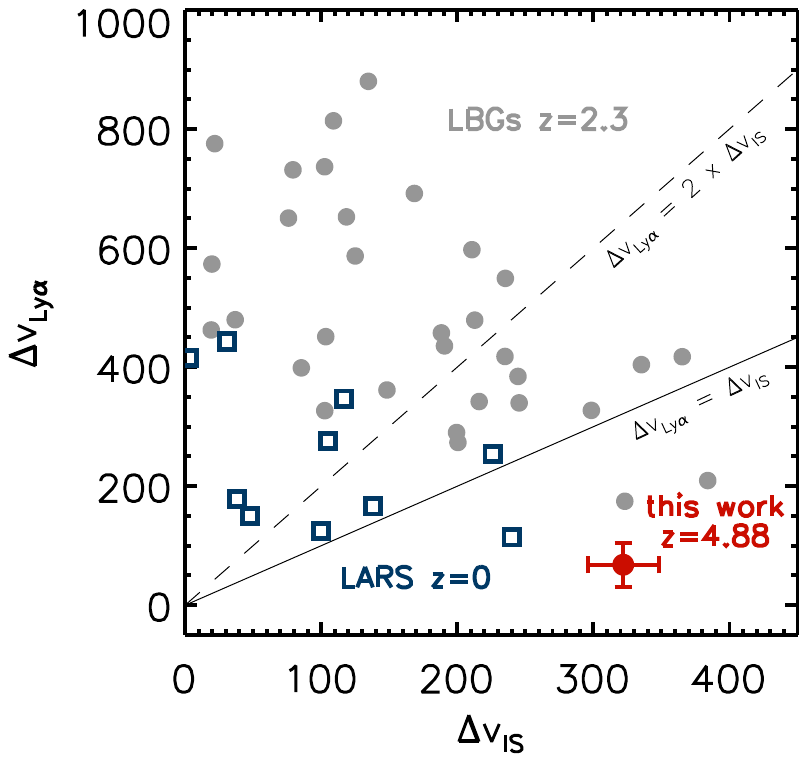}
\caption{A comparison of the velocity offsets from the systemic redshift for the Ly$\alpha$ peak and the interstellar absorption line \ion{Si}{iv} for the $z=4.88$ arc and including similar measurements at $z=0-2$ from the Ly$\alpha$ reference sample  \citep[LARS][]{Rivera2015}  and  UV-selected star-forming galaxies observed with the LRIS spectrograph on Keck \citep{Erb2006}. A ``shell''-like outflow is expected to have the Ly$\alpha$ peak be shifted by $\sim2\times \Delta v_{\rm IS}$ (dashed line), while the $z=4.88$ arc is among the very few galaxies with a Ly$\alpha$ peak shift below $\Delta v_{\rm Ly\alpha}= \Delta v _{\rm IS}$ (solid line).
}
\label{fig:RCS0224_offsets}
\end{figure}

\subsubsection{Stellar population}
\label{sec:stellar_pop}

Given  that the \ion{C}{iv} emission appears to be nebular in origin and powered by star formation (see \S\ref{sec:CIV}), we investigate the properties of the stellar population that are needed to reproduce the $\sim9${\,\AA} rest-frame equivalent width nebular \ion{C}{iv} emission. 

Figure \ref{fig:RCS0224_CIVEW} shows the evolution of the nebular \ion{C}{iv} equivalent width with metallicity, obtained from the Binary Population and Spectral Synthesis  \citep[BPASS;][]{Eldridge2012} models, using a single stellar population and including binary stellar evolution. The highly ionising photons needed to generate the high-equivalent width nebular \ion{C}{iv} lines can be generated by a young stellar population ($1-3$ Myr old) {with a low metallicity $Z=0.05Z_\odot$. 

The low-metallicity BPASS models also predict significantly reduced equivalent width stellar P-Cygni profiles, due to the fact that the winds of hot stars are driven by metal line absorption and therefore low-metallicity stars are much less efficient in driving stellar winds. This is consistent with the observed profile of \ion{C}{iv} in the $z=4.88$ arc (see Figure \ref{fig:RCS0224_CIVEW}), where we see no evidence for any redshifted stellar emission (>$500\rm\, km\,s^{-1}$ ) or broad blueshifted absorption (<$-500\rm\, km\,s^{-1}$) from the systemic velocity. In fact, a lower equivalent width stellar P-Cygni profile could provide an improved fit to our data, suggesting that the stellar iron abundance of the stellar population of the $z=4.88$ arc could be even lower than that assumed in the lowest-metallicity BPASS models available. The reasonable consistency between the stellar and nebular components in the \ion{C}{iv}  line profile provides confidence that we are indeed witnessing the early star formation in a galaxy with a very metal-poor stellar population.

\subsubsection{Kinematics}
\label{sec:kinematics}
To derive spatially resolved dynamics of the stars and gas, we exploit the enhanced spatial resolution of the strongly lensed $z=4.88$ arc and the IFU data to understand the spatial variation in the kinematics of the ISM/CGM. 

In Figure \ref{fig:RCS0224_velpos} we show the spatial variation of $\rm Ly\alpha$, \ion{C}{iv}$\lambda\lambda$1548,1551{\,\AA} and \ion{Si}{iv}$\lambda\lambda$1394,1403{\,\AA}  along the lensed galaxy image 1 running from the North-East to South-West. We use galaxy image 1, since galaxy images 2 and 3 are incomplete images that cross the critical curve (see Figure \ref{fig:RCS0224_color}) and the stellar continuum of galaxy image 4 is not spatially resolved in the MUSE data. The velocities in Figure \ref{fig:RCS0224_velpos} are given with respect to a redshift of $z_{[\ion{O}{ii}]}=4.8757$ obtained from galaxy images 2 and 3 (S07). 

As noted in \S \ref{sec:spec_lya}, the Ly$\alpha$ profile is not well described by a traditional Gaussian profile convolved with a Voigt-profile absorber and we therefore choose a non-parametric description of the Ly$\alpha$ profile. We characterise the spatial variation in the shape of the asymmetric Ly$\alpha$ profile by finding the wavelength that corresponds to 50\%, 25\% and 10\% of the peak flux redwards of the Ly$\alpha$ peak. 

The \ion{C}{iv}$\lambda\lambda$1548,1551{\,\AA} emission lines are modelled using a Gaussian emission line doublet. We model the \ion{Si}{iv}$\lambda\lambda$1394,1403{\,\AA}  absorption line doublet using Gaussians convolved with a Voigt-profile absorber. We detect these lines with $>5\sigma$ significance against the brightest continuum clump in galaxy image 1, corresponding to the North of the galaxy in the source plane. To obtain better constraints on the \ion{Si}{iv} kinematics over the whole galaxy, we combine the bright clumps in galaxy images 1, 2 and 3 that correspond to the Southern bright star-forming region in the source-plane (we also show this in Fig. \ref{fig:RCS0224_color}). 

The \ion{C}{iv} emission shows a velocity gradient of less than $50\,\rm km\,s^{-1}$ along the arc, with an irregular velocity pattern that is repeated in the lensed images 2 and 3.  This is broadly consistent with the systemic velocity derived by S07, who find a velocity gradient in [\ion{O}{ii}] of $\lesssim60\,\rm km\,s^{-1}$. Moreover, the width of the  \ion{C}{iv} doublet does not change significantly as a function of position but stays either unresolved or marginally resolved at a FWHM $\sim 100 \rm\, km\,s^{-1}$. 

 For the high-ionisation \ion{Si}{iv} line, we derive a blueshift of $300-400\,\rm km\,s^{-1}$ from the systemic redshift, consistent with the measured velocity offset by S07. The Ly$\alpha$ emission shows very little variation in the peak velocity, but broadens along the South-West side of the extended Ly$\alpha$ halo.  This is consistent with the analysis in \S\ref{sec:lya}, which suggests a collimated high-velocity outflow on top of a halo of isotropically out-flowing neutral gas. The small ($<60\,\rm km\,s^{-1}$) velocity gradient in the \ion{C}{iv} and [\ion{O}{ii}] lines as well as the $\sim100\,\rm km\,s^{-1}$ velocity gradient in the \ion{Si}{iv} absorption also support this picture of an outflow over the interpretation of large scale rotation in the halo.
 
Comparing the high-ionisation absorption features and the Ly$\alpha$ line, the Ly$\alpha$ peak produced by the receding outflow emits at significantly lower velocities ($<100\,\rm km\,s^{-1}$)  than where the absorption of  interstellar \ion{Si}{iv} takes place in the approaching outflow ($300-400\,\rm km\,s^{-1}$). This is in contradiction to a simple symmetric shell model \citep{Verhamme2006,Gronke2015}, which predicts that the Ly$\alpha$ peak be shifted by $\sim2\times \Delta v_{\rm exp}$, where $\Delta v_{\rm exp}$ is the outflow velocity of the shell as measured from the interstellar absorption features. In this model the Ly$\alpha$ peak velocity of the $z=4.88$ arc would be expected at $\sim600-800\,\rm km\,s^{-1}$, a full order of magnitude higher than our observations (see Figure \ref{fig:RCS0224_offsets}).  For comparison, the \citet{Shapley2003} composite spectrum of Lyman Break Galaxies at $z\sim3$  shows a Ly$\alpha$ velocity offset of $+360\,\rm km\,s^{-1}$ and low-ionisation lines at $-150\,\rm km\,s^{-1}$, consistent with the symmetric shell model. 

While some asymmetry could be present in the outflow, as indicated by the changing  Ly$\alpha$ linewidth on one side of the galaxy,  the peak velocity of Ly$\alpha$ changes by less than $50\,\rm km\,s^{-1}$ and it therefore seems unlikely that asymmetry in the outflowing gas explains the difference of hundreds of $\rm km\,s^{-1}$ between the approaching and receding gas tracers. We therefore suggest that a complex kinematic structure of the CGM, such as the velocity gradient we argued for in \S\ref{sec:spec_lya} must affect the absorption and escape of Ly$\alpha$ photons.

\begin{figure}
\includegraphics[scale=1.3,trim=104mm 190mm 40mm 40mm]{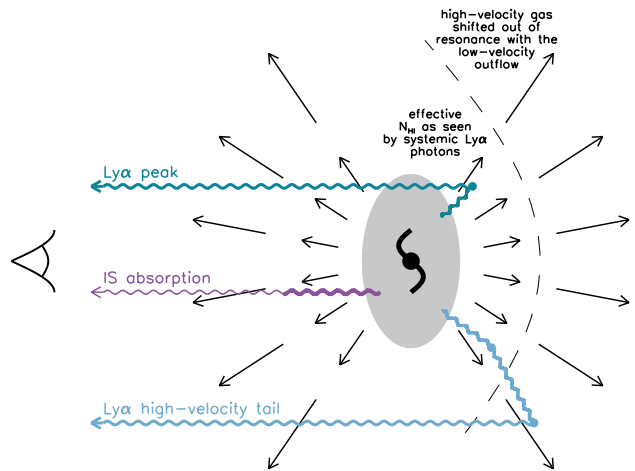}
\caption{Schematic picture of the outflows, and Ly$\alpha$ and interstellar absorption features in the $z=4.88$ arc. A strong velocity gradient in the outflow can produce the exponential Ly$\alpha$ line profile, due to the effective low column-density seen by photons emerging from the galaxy, while photons that get scattered into the higher velocity regions emerge as the lower luminosity high-velocity tail.  }
\label{fig:RCS0224_cartoon}
\end{figure}

\subsection{A physical picture for the z=4.88 arc}
\label{sec:discussion}

In \S\ref{sec:arc} we analysed the morphological and spectral properties of the Ly$\alpha$, \ion{C}{iv}, \ion{Si}{iv} and \ion{Si}{ii}  emission and absorption lines in the $z=4.88$ lensed galaxy arc behind  RCS\,0224$-$0002. Widespread nebular emission of the highly ionised \ion{C}{iv} line implies that the source is an actively star-forming galaxy with a hard ionisation field impacting upon the ISM surrounding the sites of star-formation, while the blueshifted \ion{Si}{iv}  absorption line and spatially extended redshifted Ly$\alpha$ halo indicate galaxy-wide outflows. 

A notable difficulty in this picture is the difference between the gas outflow velocities indicated by Ly$\alpha$ emission and by low-ionisation absorption, suggesting the two tracers are dominated by different parts of a CGM which hosts a complex kinematic gas structure (\S\ref{sec:kinematics}). To find a model that qualitatively describes the Ly$\alpha$ spectral properties we argued for a strong velocity gradient in the gas (\S\ref{sec:spec_lya} and Appendix \ref{app:lyaprof}), implying an accelerating outflow. In this model the velocity gradient affects the column density seen by the Ly$\alpha$ photons at any given velocity and the neutral gas at low velocity (and small radii) becomes transparent and produces the low velocity Ly$\alpha$ peak, while a small fraction of the photons is scattered to the outer, high-velocity halo (see Figure \ref{fig:RCS0224_cartoon}). This model could also explain the strong blueshift of the interstellar absorption lines compared to the peak of the Ly$\alpha$ emission, since these lines are absorbed by both low and high velocity gas.
While we expect a similar high-velocity tail in the interstellar absorption lines as is present in the Ly$\alpha$ emission line, we do not have the S/N to confirm this prediction. Finally, the spatial extend of the Ly$\alpha$ halo, which is strongly centrally peaked but shows a faint extended wing, is also well described by this model.

Accelerating outflows have already been inferred in lower redshift studies \citep[e.g.,][]{Weiner2009,Martin2009,Steidel2010}.   For example, \citet{Weiner2009} observed a ``saw-tooth'' profile and a long, high-velocity tail in the \ion{Mg}{ii} absorption features of $0.4<z<1.4$ star-forming galaxies which can be explained by accelerated cool gas.  
  \citet{Martin2009} suggest ultra-luminous infrared galaxies (ULIRGs) at $z\sim0.25$ have lower covering fractions for their higher velocity gas, implying the highest velocity gas is found at the largest radii and therefore the presence of a velocity gradient in the outflow.
Furthermore, \citet{Steidel2010} used UV-selected galaxy pairs at $z\sim2-3$ to measure the typical gas covering fraction of outflowing gas as a function of impact parameter and argued that consistency between the absorption line strength as a function of impact parameter, and the strength and profile shape of lines observed in the spectra of the galaxies, required large velocities and velocity gradients in the gas.

A physical explanation for accelerating gas is that the outflows are momentum driven \citep[e.g.,][]{Murray2005,Murray2011}. Momentum injection is  thought to be provided by radiation pressure produced on the dust grains. However, recent studies using deep ALMA observations \citep{Coppin2015,Aravena2016,Bouwens2016b,Dunlop2016} have shown that low-mass high-redshift galaxies have low dust content and the $z=4.88$ arc does not appear heavily reddened (i.e. the continuum is bluer than that observed in the composite spectrum of \citealt{Shapley2003}, see Fig. \ref{fig:RCS0224_spec}), indicating only a small fraction of the star-formation radiation is available to drive winds. \citet{Murray2005} also consider momentum injection through ram pressure by supernovae, which can deposit roughly the same amount of momentum as the radiation pressure on the dust and is therefore potentially a more likely source of momentum injection. Alternatively, \citet{Haehnelt1995} considers momentum transfer due to the radiation of ionising photons, which could be a preferred source of momentum injection given the hard ionisation radiation field we know is present throughout the galaxy, because of the wide-spread high-equivalent-width \ion{C}{iv} emission.   

In summary, a physical picture consistent with our observations is that of a vigorously star-forming galaxy, inducing a galaxy-wide momentum-driven wind, either due to supernova ram pressure or to the strong radiation field.

\subsubsection{Comparison with UV properties of sources at $z\lesssim3$}
\label{sec:comparison}
To date, only a small sample of high-redshift galaxies have been studied with high S/N rest-frame UV spectroscopy and all below $z<4$, due to their faintness and hence the long integration times needed to detect faint spectral features.  Therefore, we will compare the $z=4.88$ arc with lower redshift sources in order to understand if the features observed in this arc are common in $z\lesssim3$ galaxies, or if there is evidence for a change in the ISM/CGM properties of galaxies as we start observing sources at higher redshifts.

The brightest targets for rest-frame UV studies at $z\sim2-3$ are identified from ground-based surveys, which select strongly lensed galaxies, that are often relatively massive ($M_\ast>10^{9.5}\rm\,M_\odot$) and strongly star forming ($\rm SFR \gtrsim50\,M_\odot\,yr^{-1}$), including cB58 \citep[$z=2.73$;][]{Pettini2000,Pettini2002}, the Cosmic Eye \citep[$z=3.07$;][]{Smail2007,Quider2010}, the Cosmic Horseshoe \citep[$z=2.38$;][]{Quider2009} and SGAS J105039.6+001730 \citep[$z=3.63$;][]{Bayliss2014}. Typical UV-spectroscopic signatures in these massive galaxies include strong P-Cygni profiles seen in the \ion{C}{iv}  line profile, strong low-ionisation absorption features with respect to high-ionisation ISM lines of the same species and a wide velocity range for both low- and high-ionisation absorption lines ($\rm FWHM\sim500-1000\,km\,s^{-1}$). This is in strong contrast to the $z=4.88$ arc, where we detect no evidence for stellar winds through the  \ion{C}{iv} P-Cygni line and also where the high-ionisation ISM absorption lines are only a few hundred $\rm km\,s^{-1}$ wide, indicating a marked difference in the properties of the stellar winds and the galaxy outflows of our source. 

 A few of these massive galaxies show strong \ion{C}{iii}]$\lambda\lambda$1907,1909{\,\AA} emission, an uncommon feature in local galaxies \citep[e.g.][]{Rigby2015} as it requires a significant flux above 24 eV and therefore indicates that these high-redshift galaxies have harder ionisation fields and/or higher ionisation parameters compared to their local counterparts. However, the nebular \ion{C}{iv}$\lambda\lambda$1548,1551{\,\AA} doublet in the $z=4.88$ arc (which is only seen when significant amount of flux above 48 eV is produced) is typically not detected. 

With a dynamical mass of $\sim 10^{10}\rm\,M_\odot$ and SFR of $12\rm\,M_\odot\,\rm yr^{-1}$  (see S07) the $z=4.88$ arc behind  RCS\,0224$-$0002 might be more likely to share similar properties to lower mass sources such as the Lynx arc \citep[$z=3.36$;][]{Holden2001,Fosbury2003}, BX418 \citep[$z=2.3$;][]{Erb2010} and a sample of $z=1.4-2.9$ galaxies behind Abell 1689 and MACS 0451 targeted by \citet{Stark2014}.
Indeed, the Lynx arc and three of the 17 galaxies in the \citet{Stark2014} all show evidence for narrow \ion{C}{iv}. \citet{Erb2010} also require a significant contribution from nebular \ion{C}{iv} emission  as well as  stellar P-Cygni emisison to explain their observations. 

Due to the faintness of most of these low-mass sources a detailed analysis of the absorption features is rarely possible. \citet{Stark2014}, however, notice an almost complete absence of P-Cygni and ISM absorption features in the galaxies where they do detect the continuum (similar to local galaxies selected on their low oxygen abundance presented by \citealt{Berg2016}). Furthermore, \citet{Erb2010} are able to detect numerous absorption features in BX418 at $z=2.3$, due to an extremely deep integration, and find that the low-ionisation absorption lines in this galaxy are typically significantly weaker than the high-ionisation ISM lines, similar to the $z=4.88$ arc. An obvious difference between BX814 and the $z=4.88$ arc behind  RCS\,0224$-$0002 is the spectral shape of Ly$\alpha$; BX814 has an extremely broad, $\rm FWHM\sim850 \,km\,s^{-1}$, Ly$\alpha$ line, as opposed to $\rm FWHM<300 \,km\,s^{-1}$ observed in this galaxy. Furthermore, in BX814  the peak Ly$\alpha$ emerges at $\Delta v\rm\sim+300 \,km\,s^{-1}$, more than 3 times higher than for the $z=4.88$ arc, while the interstellar absorption lines show a $1.5-2$ times lower velocity offset ($\Delta v\rm\sim-150 \,km\,s^{-1}$).

In summary, there appear significant differences in the nebular lines, stellar and ISM absorption features between the $z=4.88$ arc behind  RCS\,0224$-$0002 and lower redshift sources with masses  above $M_\ast>10^{9.5}\rm\,M_\odot$. Typically, low-mass sources and/or low-metallicity galaxies at $z<3$ can in some cases have very similar highly ionised nebular features and some similar features in the absorption lines of low mass galaxies have been detected as well, however, no galaxy spectrum or composite spectrum of galaxies matches the full set of observations of  the $z=4.88$ arc, highlighting the need for larger samples of high S/N observations of very high-redshift galaxies in order to understand if the physical properties of the earliest systems are systematically different from their later-time counterparts or if the $z=4.88$ arc is a rare outlier in the $z\sim5$ galaxy population.

\subsubsection{Implications for reionisation studies}

Whether galaxies can reionise the Universe and what sources contribute most to reionisation, depends on a large number of parameters, including the Lyman-continuum photon production efficiency of galaxies, $\xi_{\rm ion}$, and the escape fraction of ionising photons \citep{Bouwens2015b,Robertson2015}. While determining the physical properties of galaxies in the reionisation epoch remains challenging, recent spectroscopy of $z\gtrsim7$ galaxies has shown evidence for strong rest-frame UV nebular emission lines such as \ion{C}{iv}$\lambda\lambda$1548,1551{\,\AA} \citep{Stark2015,Stark2017}. \textit{Spitzer}/IRAC imaging studies have also inferred extremely strong  [\ion{O}{iii}]$\lambda\lambda$4959,5007{\,\AA} in the rest-frame optical spectra of typical $z\sim7-8$ galaxies \citep{Labbe2013,Smit2014,Smit2015,RobertsBorsani2016}. These results suggest that galaxies in the reionisation epoch could have similar hard radiation field and/or high ionisation parameter as the $z=4.88$ arc and it is therefore interesting to assess this galaxy as an analogue of the sources that might be responsible for reionisation. 

Using the BPASS stellar population template of a young extremely low metallicity galaxy needed to match the \ion{C}{iv} equivalent width (see \S\ref{sec:stellar_pop}) we derive a Lyman-continuum photon production efficiency in the $z=4.88$ arc of  $\log_{10}\xi_{\rm ion}=25.74 \,\rm Hz\,erg^{-1}$, 0.63 dex higher than the canonical value of $\log_{10}\xi_{\rm ion}=25.11 \,\rm Hz\,erg^{-1}$ \citep{Kennicutt1998}. Systemic deviations of $\sim0.1$ dex from the canonical value of $\xi_{\rm ion}$ have been derived from the inferred H$\alpha$ emission in typical $z\sim4-5$ UV-selected galaxies \citep{Bouwens2016,Smit2016,Rasappu2016}.  The bluest $z\sim4-5$ galaxies ($\beta$<-2.3), likely very young and dust-free sources, have a significantly higher Lyman-continuum photon production efficiency  of $\log_{10}\xi_{\rm ion}=25.53-25.78 \,\rm Hz\,erg^{-1}$ \citep{Bouwens2016,Stark2015}, in good agreement with our derived value for the $z=4.88$ arc, which also has a blue UV-continuum colour ($\beta$=-2.2). If similarly young  and low-metallicity galaxies are common at $z\gtrsim7$, they could contribute significantly to reionisation even for modest ($\lesssim5$\%) escape fractions. 

Measuring the direct escape of Lyman-continuum photons at $z\sim5$ is challenging due to the intervening Ly$\alpha$ forest in the IGM. However, the \ion{Si}{ii}$\lambda$1304{\,\AA} absoprtion in the $z=4.88$ arc shows no sign of absorption at the systemic velocity by low-ionisation gas in the ISM, indicating Lyman-continuum photons might also easily escape. Furthermore, the collimated high-velocity outflow discussed in \S\ref{sec:spec_lya} could blow holes into the ISM through which the photons preferentially escape \citep[see][]{Erb2015}.   

Another important implication of the $z=4.88$ arc as an $z\gtrsim7$ galaxy analogue, is the opportunity to identify similar sources with future facilities such as the \textit{James Webb Space Telescope (JWST)} and the various Extremely Large Telescopes (ELTs). Most of the Ly$\alpha$ emission of galaxies in the reionisation epoch will be absorbed due to the surrounding neutral IGM, but strong nebular lines such as \ion{C}{iv}$\lambda\lambda$1548,1551{\,\AA} (seen at an observed EW of $\sim80\,${\,\AA} at $z\sim8$) can be easily identified. While these lines are uncommon in the local Universe, our results indicate that these lines are widely produced by the young and low-metallicity stellar population within the $z=4.88$ arc. These characteristics are likely to be more common as we start observing galaxies at earlier epochs.

\begin{figure}
\includegraphics[scale=1.,trim=25mm 185mm 100mm 35mm]{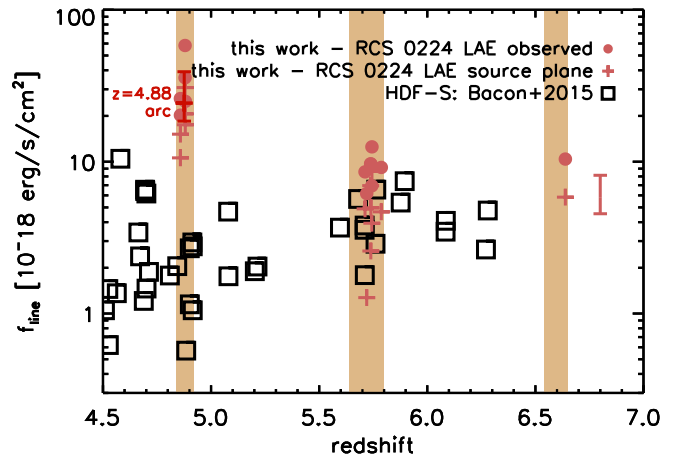}
\caption{ A comparison of observed and intrinsic flux of Ly$\alpha$ emitters in the HDF-S from \citet{Bacon2015} with our blind search in the RCS\,0224$-$0002 MUSE field. The typical uncertainty on the intrinsic fluxes is indicated by the errorbars on the right. The shaded regions indicate our search windows, which have low sky contamination. Comparing our 3.75-hour integration to the 27-hour MUSE integration by \citet{Bacon2015} we are able to detect similarly faint sources in the source-plane, that is amplified by the cluster lensing.
 }
\label{fig:RCS0224_Lyacomparison}
\end{figure}

\section{A Blind search for high-redshift Ly${\alpha}$ emitters}
\label{sec:blind_search}
Finally, we can use the relatively wide field of view of MUSE to search for other emitters in the field. 
Deep MUSE observations in the $Hubble$ deep fields have proven efficient in detecting Ly$\alpha$ out to redshift $z\sim6.6$ \citep[e.g.,][]{Bacon2015}. The extremely faint or undetected $HST$ counterparts of these high-redshift Ly$\alpha$ emitters indicate we are sensitive to the faint end of the UV luminosity function; sources with similar properties to the galaxies expected to be responsible for cosmic reionisation. While these observations require integration times of $\sim30$ hours, similar sources can potentially be found behind by strong lensing clusters within reasonable integrations times \citep[e.g.,][]{Karman2015,Bina2016}. 


Due to its high mass, relative compactness and high redshift RCS\,0224$-$0002 appears to be an efficient lens for high-redshift galaxies. Unlike low-redshift strong-lensing clusters, our single MUSE pointing covers the entire $z\sim6$ critical curves of RCS\,0224$-$0002. We therefore explore the potential of RCS\,0224$-$0002 as a window into the very high-redshift Universe.  We search for emission line candidates in our MUSE dataset in three windows with low sky contamination, 7100-7200{\,\AA}, 8070-8270{\,\AA} and 9060-9300{\,\AA}, corresponding to Ly$\alpha$ redshift ranges of $z=4.84-4.92$, $z=5.64-5.80$ and $z=6.54-6.65$ respectively. To achieve this, we develop a blind line detection method that follows a number of consecutive steps to identify extremely faint sources, while minimizing spurious detections. First, for each pixel in the MUSE field of view (masking bright continuum sources and removing cluster galaxies from the sample), we extract a one-dimensional spectrum by averaging $5\times5$ pixels, where we use the PSF measured from the MUSE data to assign a weight to each of the pixels. For each of the one-dimensional spectra we search for individual spectral pixels that have a value $>3.5\sigma$ above the noise (estimated from the same one-dimensional spectrum, but masking the skylines) to identify potential lines. For each candidate, we fit a single Gaussian profile to the spectrum and we require a $\Delta\chi^2\geq7.5^2$ between the Gaussian and a straight line fit with a constant value. 

To remove spurious detections we next generate a $12\times12$ arcsec continuum-subtracted narrowband image around the line. We create this image by extracting the images from the cube with a wavelength within the FWHM of the line (as measured from the Gaussian profile fit) and averaging them, we then subtract the median continuum in two bands (20{\,\AA} wide) at each side of the line (20{\,\AA} removed from the centre). On the continuum-subtracted narrowband image we measure 1000 randomly selected point-source fluxes, using the PSF extracted from a isolated star in the MUSE continuum image to weight the pixels around each point. We require the line candidates to have a flux $\geq3\sigma$ above the random sampled point-source flux distribution. 

We visually inspect our line-candidates and remove all sources that are clearly [\ion{O}{ii}], [\ion{O}{iii}] or H$\alpha$ emission and we remove low-redshift interlopers when bright continuum flux is detected blueward of the rest-frame 912{\,\AA} limit (based on the Ly$\alpha$ redshift). Note that while we do not exclude sources which show continuum flux between rest-frame 912{\,\AA} and 1216{\,\AA}, we detect flux below the rest-frame 1216{\,\AA} break in only one Ly$\alpha$ candidate (RCS0224\_LAEz4p8773) from our final sample. Furthermore, we remove sources that appear due to noise artefacts at the edges of the CCDs, or sources that are strongly affected by an uneven background. 

A significant uncertainty in our interloper rejection is that almost none of our Ly$\alpha$ candidates are detected in the $HST$/ACS+WFC3 imaging, and therefore no prior based on photometric redshift of the sources can be applied to our Ly$\alpha$ selection. Furthermore, many of our Ly$\alpha$ candidates are observed with too low S/N to identify the expected asymmetric Ly$\alpha$ spectral line shape and therefore our sample could be contaminated by for example high-equivalent-width [\ion{O}{iii}]$\lambda$5007{\,\AA} lines of lower redshift galaxies, for which the [\ion{O}{iii}]$\lambda$4959{\,\AA} line is too faint to be detected. 

To make an estimate of the number of spurious detections we expect in our blind search we perform a test for false-positives by running our source selection code on the inverted data-cube. To identify a ``pure'' sample, we calculate the S/N required in the algorithm to give a false positive rate of zero; which is $\Delta\chi^2\geq7.5^2$ for the line fit and $\geq3\sigma$ for the continuum-subtracted point source flux. 

In our final sample of Ly$\alpha$ candidates we have five sources at $z\sim4.8$, eight sources at $z\sim5.7$ and one source at $z\sim6.6$. The sources are listed in Table \ref{tab:lya} and thumbnails of all the sources presented in Appendix \ref{app:lyaem}. For comparison, in a 30h exposure over a 1 arcmin$^2$ field over the HDF-S \citet{Bacon2015} find for the same redshift intervals seven Ly$\alpha$ emitters at $z\sim4.8$, six emitters at $z\sim5.7$ and no emitters at $z\sim6.6$, see Figure \ref{fig:RCS0224_Lyacomparison}. Using a 4h exposure and 1 arcmin$^2$ field, \citet{Karman2015} find only 1 (multiply-lensed) source above $z>4.5$ behind the Frontier Fields cluster Abell S1063.

A notable result from our blind search for Ly$\alpha$ line candidates  is the presence of a large number of bright Ly$\alpha$ emitting sources at  $z\sim4.88$. Three bright sources at  $\rm 2.5-\rm 5.8\times10^{-17}\,erg\,s^{-1}\,cm^{-2}$ are located at $\sim+250-300\rm\,km\,s^{-1}$ or $\sim200$ kpc from the bright arc, while two more sources are located $\sim-1000\rm\,km\,s^{-1}$ or $\sim1.7$ Mpc from the arc. With the exception of one of these sources, none of the line-emitters are detected in the $HST$ imaging (at $>5\sigma$), however they all exhibit a clear asymmetric line profile that provides evidence of these sources being Ly$\alpha$ emitters. After correcting for the lensing magnification these sources have Ly$\alpha$ line luminosities $\sim1-3 \times 10^{-17}\,\rm\,erg\,s^{-1}\,cm^{-2}$ ($L=2.6-5.1\times 10^{42}\,\rm\,erg\,s^{-1}$), $\sim2-10\times$  brighter than sources found in the $Hubble$ Deep Field South by \citet{Bacon2015} at the same redshift. Comparing to the faint end of the Ly$\alpha$ luminosity functions obtained through narrow-band surveys at $z=3.5-5.7$ \citep[e.g.][]{Ouchi2008,Santos2016}, we would predict $0.1-0.9$ Ly$\alpha$ emitters as bright as $\log{L/\rm erg\,s^{-1}}=42.5$ in our MUSE data in the redshift window $z=4.84-4.92$.
This suggests that either  the Ly$\alpha$ luminosity function is steeper at the faint-end than measured in narrow-band surveys \citep[see also][]{Drake2016}, or else that the $z=4.88$ arc is located in a $\sim 7-60\times$ over-dense region, or group.

\section{Summary}
\label{sec:summary}
We present a survey for line emitter galaxies behind the strong lensing cluster RCS\,0224$-$0002. We analyse the rest-frame UV spectrum of a lensed galaxy  magnified 29 times at $z=4.88$. For this source we observe the following properties:

\begin{itemize}
\item The $z=4.88$ galaxy is surrounded by a spatially extended Ly$\alpha$ halo with an exponential spatial profile. The spectral properties of the Ly$\alpha$ halo are spatially-uniform, showing a single redshifted peak close to the systemic velocity ($\Delta v<100\rm\,km\,s^{-1}$) and  high-velocity tail ($\Delta v>1000\rm\,km\,s^{-1}$). The spatial and spectral properties of the halo are consistent with resonantly scattered Ly$\alpha$ photons produced in a central source and backscattered in a receding outflow from the galaxy. 

\item We detect spatially resolved narrow \ion{C}{iv}$\lambda\lambda$1548,1551{\,\AA} emission. The spatial distribution of \ion{C}{iv} strongly resembles that of the [\ion{O}{ii}] line, suggesting a nebular origin of the line, powered by star-formation. We argue that the strong \ion{C}{iv} emission ($\rm EW_0\sim9$\AA) can be reproduced with a young ($t<5\rm\,Myr$),  low-metallicity ($Z\lesssim0.05Z_\odot$) stellar population. The blue UV-continuum color ($\beta$=-2.2) and the absence of a P-Cygni profile, indicating low-metallicity stars with significantly reduced stellar winds, is consistent with this analysis.  

\item We observe strong high-ionisation interstellar absorption lines in \ion{C}{iv} and \ion{S}{iv} with a significant blueshift ($\Delta v\sim300\rm\,km\,s^{-1}$) from the systemic velocity and much weaker low-ionisation \ion{S}{ii} absorption (EW(\ion{Si}{ii}$\lambda$1304{\,\AA})/EW(\ion{Si}{iv}$\lambda$1394{\,\AA})=0.2). The blueshift of the interstellar lines is surprising when considering how close to the systemic velocity we observe the Ly$\alpha$ line and given that an outflowing-shell model suggests $\Delta v_{\rm IS}\sim v_{\rm shell}$ and $\Delta v_{\rm Ly\alpha}\sim 2\times v_{\rm shell}$. 
\end{itemize}
We propose a physical model for this galaxy in which the outflowing gas follows a strong velocity gradient such that the effective column density of neutral gas, as seen by the outwards scattering Ly$\alpha$ photons, is significantly reduced, allowing for Ly$\alpha$ to escape at much lower velocities than the mean gas outflow \citep[see][]{Verhamme2006}. This velocity gradient likely requires a momentum ejection into the gas, which can originate from supernovae ram pressure or radiation pressure \citep[e.g.][]{Murray2005}. These results emphasize the importance of increasing the samples of high-redshift low-mass galaxies where we are able to detect the interstellar absorption features, as relying on Ly$\alpha$ as a tracer of galaxy outflows can significantly underestimate the feedback in galaxies such as the $z=4.88$ arc behind RCS\,0224.  

We perform a blind line search for high-redshift Ly$\alpha$ using three wavelength ranges that are relatively free of sky lines, corresponding to $z_{\rm Ly\alpha}=4.84-4.92$, $z_{\rm Ly\alpha}=5.64-5.80$ and $z_{\rm Ly\alpha}=6.54-6.65$. We select sources above the significance level needed such that a line search on the inverted data results in zero false positives. We find a total of 14 Ly$\alpha$ candidates, of which only one is detected in the $HST$ imaging. This suggests line surveys over strong lensing clusters with MUSE are efficient at finding ultra-faint galaxies out to $z\sim6.6$ and hence study the properties of faint Ly$\alpha$ emitting galaxies that are likely to have contributed to reionisation.

\section*{Acknowledgments}
We are grateful to Graham Smith for recovering the parameters of the S07 lensing model. We thank Matthew Hayes, Bethan James, Vera Patricio, Max Pettini, Tom Theuns, Ryan Trainor and Anne Verhamme for useful discussions. We are grateful to Max Gronke for giving us access to the on-line tool {\small TLAC\_WEB}. RS, AMS, RJM and IRS acknowledge support from STFC (ST/L0075X/1). RS and IRS also acknowledge support from the ERC Advanced Investigator programme DUSTYGAL 321334.  In addition, RS acknowledges support from the Leverhulme Trust, AMS from an STFC Advanced Fellowship (ST/H005234/1), RJM acknowledges support from a Royal Society URF and IRS acknowledges support from a Royal Society/Wolfson Merit Award. JPK acknowledges support from the ERC advanced grant LIDA and from CNRS.

\bibliographystyle{plainnat}

\begin{thebibliography}{}
\bibitem[Aravena et al.(2016)]{Aravena2016} Aravena, M., Decarli, R., Walter, F., et al.\ 2016, \apj, 833, 68 
\bibitem[Bacon et al.(2010)]{Bacon2010} Bacon, R., Accardo, M., Adjali, L., et al.\ 2010, \procspie, 7735, 773508 
\bibitem[Bacon et al.(2015)]{Bacon2015} Bacon, R., Brinchmann, J., Richard, J., et al.\ 2015, \aap, 575, A75 
\bibitem[Bayliss et al.(2014)]{Bayliss2014} Bayliss, M.~B., Rigby, J.~R., Sharon, K., et al.\ 2014, \apj, 790, 144 
\bibitem[Berg et al.(2016)]{Berg2016} Berg, D.~A., Skillman, E.~D., Henry, R.~B.~C., Erb, D.~K., \& Carigi, L.\ 2016, \apj, 827, 126 
\bibitem[Bina et al.(2016)]{Bina2016} Bina, D., Pell{\'o}, R., Richard, J., et al.\ 2016, \aap, 590, A14 
\bibitem[Bouwens et al.(2015a)]{Bouwens2015} Bouwens, R.~J., Illingworth, G.~D., Oesch, P.~A., et al.\ 2015a, \apj, 803, 34
\bibitem[Bouwens et al.(2015b)]{Bouwens2015b} Bouwens, R.~J., Illingworth, G.~D., Oesch, P.~A., et al.\ 2015b, \apj, 811, 140
\bibitem[Bouwens et al.(2016)]{Bouwens2016} Bouwens, R.~J., Smit, R., Labb{\'e}, I., et al.\ 2016, \apj, 831, 176 
\bibitem[Bouwens et al.(2016)]{Bouwens2016b} Bouwens, R.~J., Aravena, M., Decarli, R., et al.\ 2016, \apj, 833, 72 
\bibitem[Bowler et al.(2015)]{Bowler2015} Bowler, R.~A.~A., Dunlop, J.~S., McLure, R.~J., et al.\ 2015, \mnras, 452, 1817 
\bibitem[Caminha et al.(2016)]{Caminha2016} Caminha, G.~B., Karman, W., Rosati, P., et al.\ 2016, \aap, 595, A100 
\bibitem[Christensen et al.(2012a)]{Christensen2012a} Christensen, L., Richard, J., Hjorth, J., et al.\ 2012a, \mnras, 427, 1953 
\bibitem[Christensen et al.(2012b)]{Christensen2012} Christensen, L., Laursen, P., Richard, J., et al.\ 2012b, \mnras, 427, 1973 
\bibitem[Coppin et al.(2015)]{Coppin2015} Coppin, K.~E.~K., Geach, J.~E., Almaini, O., et al.\ 2015, \mnras, 446, 1293 
\bibitem[Dessauges-Zavadsky et al.(2010)]{Dessauges2010} Dessauges-Zavadsky, M., D'Odorico, S., Schaerer, D., et al.\ 2010, \aap, 510, A26 
\bibitem[Dijkstra et al.(2006)]{Dijkstra2006} Dijkstra, M., Haiman, Z., \& Spaans, M.\ 2006, \apj, 649, 14 
\bibitem[Dijkstra \& Loeb(2009)]{Dijkstra2009} Dijkstra, M., \& Loeb, A.\ 2009, \mnras, 400, 1109
\bibitem[Drake et al.(2016)]{Drake2016} Drake, A.~B., Guiderdoni, B., Blaizot, J., et al.\ 2016, arXiv:1609.02920 
\bibitem[Dunlop et al.(2016)]{Dunlop2016} Dunlop, J.~S., McLure, R.~J., Biggs, A.~D., et al.\ 2016, arXiv:1606.00227 
\bibitem[Eldridge \& Stanway(2012)]{Eldridge2012} Eldridge, J.~J., \& Stanway, E.~R.\ 2012, \mnras, 419, 479 
\bibitem[Erb et al.(2006)]{Erb2006} Erb, D.~K., Steidel, C.~C., Shapley, A.~E., et al.\ 2006, \apj, 646, 107 
\bibitem[Erb et al.(2010)]{Erb2010} Erb, D.~K., Pettini, M., Shapley, A.~E., et al.\ 2010, \apj, 719, 1168 
\bibitem[Erb(2015)]{Erb2015} Erb, D.~K.\ 2015, \nat, 523, 169 
\bibitem[Faucher-Gigu{\`e}re et al.(2010)]{Faucher2010} Faucher-Gigu{\`e}re, C.-A., Kere{\v s}, D., Dijkstra, M., Hernquist, L., \& Zaldarriaga, M.\ 2010, \apj, 725, 633 
\bibitem[Feltre et al.(2016)]{Feltre2016} Feltre, A., Charlot, S., \& Gutkin, J.\ 2016, \mnras, 456, 3354 
\bibitem[Finkelstein et al.(2015)]{Finkelstein2015} Finkelstein, S.~L., Ryan, R.~E., Jr., Papovich, C., et al.\ 2015, \apj, 810, 71 
\bibitem[Fosbury et al.(2003)]{Fosbury2003} Fosbury, R.~A.~E., Villar-Mart{\'{\i}}n, M., Humphrey, A., et al.\ 2003, \apj, 596, 797 
\bibitem[Giallongo et al.(2015)]{Giallongo2015} Giallongo, E., Grazian, A., Fiore, F., et al.\ 2015, \aap, 578, A83 
\bibitem[Giavalisco et al.(2004)]{Giavalisco2004} Giavalisco, M., Dickinson, M., Ferguson, H.~C., et al.\ 2004, \apjl, 600, L103 
\bibitem[Gladders et al.(2002)]{Gladders2002} Gladders, M.~D., Yee, H.~K.~C., \& Ellingson, E.\ 2002, \aj, 123, 1 
\bibitem[Gronke et al.(2015)]{Gronke2015} Gronke, M., Bull, P., \& Dijkstra, M.\ 2015, \apj, 812, 123 
\bibitem[Gullberg et al.(2016)]{Gullberg2016} Gullberg, B., De Breuck, C., Lehnert, M.~D., et al.\ 2016, \aap, 586, A124 
\bibitem[Haehnelt(1995)]{Haehnelt1995} Haehnelt, M.~G.\ 1995, \mnras, 273, 249 
\bibitem[Hainline et al.(2011)]{Hainline2011} Hainline, K.~N., Shapley, A.~E., Greene, J.~E., \& Steidel, C.~C.\ 2011, \apj, 733, 31
\bibitem[Hayes et al.(2013)]{Hayes2013} Hayes, M., {\"O}stlin, G., Schaerer, D., et al.\ 2013, \apjl, 765, L27 
\bibitem[Holden et al.(2001)]{Holden2001} Holden, B.~P., Stanford, S.~A., Rosati, P., et al.\ 2001, \aj, 122, 629
\bibitem[Jones et al.(2012)]{Jones2012} Jones, T., Stark, D.~P., \& Ellis, R.~S.\ 2012, \apj, 751, 51 
\bibitem[Jullo et al.(2007)]{Jullo2007} Jullo, E., Kneib, J.-P., Limousin, M., et al.\ 2007, New Journal of Physics, 9, 447  
\bibitem[Jullo \& Kneib(2009)]{Jullo2009} Jullo, E., \& Kneib, J.-P.\ 2009, \mnras, 395, 1319 
\bibitem[Karman et al.(2015)]{Karman2015} Karman, W., Caputi, K.~I., Grillo, C., et al.\ 2015, \aap, 574, A11 
\bibitem[Kennicutt(1998)]{Kennicutt1998} Kennicutt, R.~C., Jr.\ 1998, \araa, 36, 189 
\bibitem[Kneib et al.(1996)]{Kneib1996} Kneib, J.-P., Ellis, R.~S., Smail, I., Couch, W.~J., \& Sharples, R.~M.\ 1996, \apj, 471, 643 
\bibitem[Labb{\'e} et al.(2013)]{Labbe2013} Labb{\'e}, I., Oesch, P.~A., Bouwens, R.~J., et al.\ 2013, \apjl, 777, L19 
\bibitem[Lehnert \& Bremer(2003)]{Lehnert2003} Lehnert, M.~D., \& Bremer, M.\ 2003, \apj, 593, 630 
\bibitem[Leitherer et al.(2001)]{Leitherer2001} Leitherer, C., Le{\~a}o, J.~R.~S., Heckman, T.~M., et al.\ 2001, \apj, 550, 724 
\bibitem[Leitherer et al.(2011)]{Leitherer2011} Leitherer, C., Tremonti, C.~A., Heckman, T.~M., \& Calzetti, D.\ 2011, \aj, 141, 37 
\bibitem[Matsuda et al.(2012)]{Matsuda2012} Matsuda, Y., Yamada, T., Hayashino, T., et al.\ 2012, \mnras, 425, 878 
\bibitem[Meneghetti et al.(2016)]{Meneghetti2016} Meneghetti, M., Natarajan, P., Coe, D., et al.\ 2016, arXiv:1606.04548
\bibitem[Momose et al.(2014)]{Momose2014} Momose, R., Ouchi, M., Nakajima, K., et al.\ 2014, \mnras, 442, 110 
\bibitem[Madau et al.(1996)]{Madau1996} Madau, P., Ferguson, H.~C., Dickinson, M.~E., et al.\ 1996, \mnras, 283, 1388 
\bibitem[Martin \& Bouch{\'e}(2009)]{Martin2009} Martin, C.~L., \& Bouch{\'e}, N.\ 2009, \apj, 703, 1394 
\bibitem[McLure et al.(2009)]{Mclure2009} McLure, R.~J., Cirasuolo, M., Dunlop, J.~S., Foucaud, S., \& Almaini, O.\ 2009, \mnras, 395, 2196 
\bibitem[Murray et al.(2005)]{Murray2005} Murray, N., Quataert, E., \& Thompson, T.~A.\ 2005, \apj, 618, 569 
\bibitem[Murray et al.(2011)]{Murray2011} Murray, N., M{\'e}nard, B., \& Thompson, T.~A.\ 2011, \apj, 735, 66 
\bibitem[Oke \& Gunn(1983)]{OkeGun} Oke, J.~B., \& Gunn, J.~E.\ 1983, \apj, 266, 713
\bibitem[Ouchi et al.(2004)]{Ouchi2004} Ouchi, M., Shimasaku, K., Okamura, S., et al.\ 2004, \apj, 611, 660 
\bibitem[Ouchi et al.(2008)]{Ouchi2008} Ouchi, M., Shimasaku, K., Akiyama, M., et al.\ 2008, \apjs, 176, 301-330 
\bibitem[Heckman et al.(2015)]{Heckman2015} Heckman, T.~M., Alexandroff, R.~M., Borthakur, S., Overzier, R., \& Leitherer, C.\ 2015, \apj, 809, 147  
\bibitem[Patr{\'{\i}}cio et al.(2016)]{Patricio2016} Patr{\'{\i}}cio, V., Richard, J., Verhamme, A., et al.\ 2016, \mnras, 456, 4191 
\bibitem[Petrosian(1976)]{Petrosian1976} Petrosian, V.\ 1976, \apjl, 209, L1 
\bibitem[Pettini et al.(2000)]{Pettini2000} Pettini, M., Steidel, C.~C., Adelberger, K.~L., Dickinson, M., \& Giavalisco, M.\ 2000, \apj, 528, 96 
\bibitem[Pettini et al.(2002)]{Pettini2002} Pettini, M., Rix, S.~A., Steidel, C.~C., et al.\ 2002, \apj, 569, 742 
\bibitem[Quider et al.(2009)]{Quider2009} Quider, A.~M., Pettini, M., Shapley, A.~E., \& Steidel, C.~C.\ 2009, \mnras, 398, 1263 
\bibitem[Quider et al.(2010)]{Quider2010} Quider, A.~M., Shapley, A.~E., Pettini, M., Steidel, C.~C., \& Stark, D.~P.\ 2010, \mnras, 402, 1467 
\bibitem[Rasappu et al.(2016)]{Rasappu2016} Rasappu, N., Smit, R., Labb{\'e}, I., et al.\ 2016, \mnras, 461, 3886 
\bibitem[Rigby et al.(2015)]{Rigby2015} Rigby, J.~R., Bayliss, M.~B., Gladders, M.~D., et al.\ 2015, \apjl, 814, L6
\bibitem[Rivera-Thorsen et al.(2015)]{Rivera2015} Rivera-Thorsen, T.~E., Hayes, M., {\"O}stlin, G., et al.\ 2015, \apj, 805, 14 
\bibitem[Roberts-Borsani et al.(2016)]{RobertsBorsani2016} Roberts-Borsani, G.~W., Bouwens, R.~J., Oesch, P.~A., et al.\ 2016, \apj, 823, 143 
\bibitem[Robertson et al.(2015)]{Robertson2015} Robertson, B.~E., Ellis, R.~S., Furlanetto, S.~R., \& Dunlop, J.~S.\ 2015, \apjl, 802, L19  
\bibitem[Rosdahl \& Blaizot(2012)]{Rosdahl2012} Rosdahl, J., \& Blaizot, J.\ 2012, \mnras, 423, 344 
\bibitem[Sawicki et al.(1997)]{Sawicki1997} Sawicki, M.~J., Lin, H., \& Yee, H.~K.~C.\ 1997, \aj, 113, 1 
\bibitem[Santos et al.(2016)]{Santos2016} Santos, S., Sobral, D., \& Matthee, J.\ 2016, \mnras,  
\bibitem[Shapley et al.(2003)]{Shapley2003} Shapley, A.~E., Steidel, C.~C., Pettini, M., \& Adelberger, K.~L.\ 2003, \apj, 588, 65 
\bibitem[Smail et al.(2007)]{Smail2007} Smail, I., Swinbank, A.~M., Richard, J., et al.\ 2007, \apjl, 654, L33 
\bibitem[Smit et al.(2014)]{Smit2014} Smit, R., Bouwens, R.~J., Labb{\'e}, I., et al.\ 2014, \apj, 784, 58 
\bibitem[Smit et al.(2015)]{Smit2015} Smit, R., Bouwens, R.~J., Franx, M., et al.\ 2015, \apj, 801, 122 
\bibitem[Smit et al.(2016)]{Smit2016} Smit, R., Bouwens, R.~J., Labb{\'e}, I., et al.\ 2016, \apj, 833, 254 
\bibitem[Stark et al.(2014)]{Stark2014} Stark, D.~P., Richard, J., Siana, B., et al.\ 2014, \mnras, 445, 3200 
\bibitem[Stark et al.(2015)]{Stark2015} Stark, D.~P., Walth, G., Charlot, S., et al.\ 2015, \mnras, 454, 1393 
\bibitem[Stark et al.(2017)]{Stark2017} Stark, D.~P., Ellis, R.~S., Charlot, S., et al.\ 2017, \mnras, 464, 469 
\bibitem[Steidel et al.(1996)]{Steidel1996} Steidel, C.~C., Giavalisco, M., Pettini, M., Dickinson, M., \& Adelberger, K.~L.\ 1996, \apjl, 462, L17 
\bibitem[Steidel et al.(1999)]{Steidel1999} Steidel, C.~C., Adelberger, K.~L., Giavalisco, M., Dickinson, M., \& Pettini, M.\ 1999, \apj, 519, 1 
\bibitem[Steidel et al.(2010)]{Steidel2010} Steidel, C.~C., Erb, D.~K., Shapley, A.~E., et al.\ 2010, \apj, 717, 289 
\bibitem[Steidel et al.(2011)]{Steidel2011} Steidel, C.~C., Bogosavljevi{\'c}, M., Shapley, A.~E., et al.\ 2011, \apj, 736, 160 
\bibitem[Swinbank et al.(2007)]{Swinbank2007} Swinbank, A.~M., Bower, R.~G., Smith, G.~P., et al.\ 2007, \mnras, 376, 479 
\bibitem[Swinbank et al.(2015)]{Swinbank2015} Swinbank, A.~M., Vernet, J.~D.~R., Smail, I., et al.\ 2015, \mnras, 449, 1298 
\bibitem[van der Burg et al.(2010)]{vanderBurg2010} van der Burg, R.~F.~J., Hildebrandt, H., \& Erben, T.\ 2010, \aap, 523, A74 
\bibitem[Vanzella et al.(2016)]{Vanzella2016} Vanzella, E., De Barros, S., Cupani, G., et al.\ 2016, \apjl, 821, L27 
\bibitem[Vanzella et al.(2017)]{Vanzella2017} Vanzella, E., Balestra, I., Gronke, M., et al.\ 2017, \mnras, 465, 3803 
\bibitem[Verhamme et al.(2006)]{Verhamme2006} Verhamme, A., Schaerer, D., \& Maselli, A.\ 2006, \aap, 460, 397 
\bibitem[Weiner et al.(2009)]{Weiner2009} Weiner, B.~J., Coil, A.~L., Prochaska, J.~X., et al.\ 2009, \apj, 692, 187 
\bibitem[Wisotzki et al.(2016)]{Wisotzki2016} Wisotzki, L., Bacon, R., Blaizot, J., et al.\ 2016, \aap, 587, A98 


\end{thebibliography}


\appendix
\section{Best-fit parameters of the lens model}
\label{app:table_lensmodel}
In \S\ref{sec:lensmodel} we described the set-up and main results of our {\small LENSTOOL} modelling. In this appendix we present the full observational constraints in Table \ref{tab:lens_constraints} and the best fit parameters of the lensing model that is used in this work in Table \ref{tab:lensparam}.

  \begin{table}
\centering
\caption{Input locations of strong lens multiple images [RA, Dec], uncertainty on the position [arcsec], redshift. The redshift of system D is left free, and constrained during the fit to be >5.3 (1sigma).}
\begin{tabular}{lcccc}
\hline
\hline
ID & RA & Dec &  $\Delta$[$\arcsec$]  & $z$ \\
\hline

A1 & 36.140826   & -0.038235783   & 0.2   & 4.88      \\
A2 & 36.139833   &  -0.039005229  &  0.2  &  4.88     \\
A3 & 36.138742   & -0.040753058   & 0.2   & 4.88      \\
A4 & 36.14349    & -0.042785468   & 0.2   & 4.88      \\
B1 & 36.144748   &  -0.042371596  &  0.2  &  2.395    \\
B2 & 36.145091   &   -0.041225429 &  0.2  &  2.395    \\
B3 & 36.143675   &  -0.039536976  &   0.8 &  2.395    \\
B4 & 36.141663   &  -0.039894484  &  0.8  &  2.395    \\
B5 & 36.140316   &  -0.043556393  &  0.8  &  2.395    \\
B6 & 36.142475   &  -0.041860145  &  0.2  &  2.395    \\
C1 &  36.14525   &   -0.03783889  &  0.8  &  5.498    \\
C2 &  36.141739  & -0.043416674   & 0.2   & 5.498     \\
D1 &   36.13902  & -0.043164193   & 0.8   & -         \\
D2 &   36.142798 &  -0.03823784   & 0.8   & -        \\
D3 &   36.140498 &  -0.03920681   & 0.8   & -		  \\
\hline

\end{tabular}
\label{tab:lens_constraints}
\end{table}

  \begin{table}
\centering
\caption{Parameters of best-fitting mass model. Angles of the major axis are anticlockwise from North. }
\begin{tabular}{lcccccc}
\hline
\hline
$\Delta$RA & $\Delta$Dec & $v_\mathrm{disp}$ & $\varepsilon$ & $\theta$ & $r_\mathrm{core}$  & $r_\mathrm{cut}$  \\
 { [$\arcsec$]} &  [$\arcsec$] & [km/s] &  &  [deg] & [kpc] &  [kpc] \\
\hline
$ -1.80$ & $  3.80$ & $624.7$ & $  0.57$ & $128$ & $ 40.01$ & $1000.0$ \\
$  0.00$ & $  0.00$ & $385.7$ & $  0.57$ & $ 11$ & $ 14.61$ & $1000.0$ \\
$ -1.29$ & $  2.23$ & $190.3$ & $  0.16$ & $ 29$ & $  0.51$ & $  77.4$ \\
$  4.16$ & $  5.43$ & $186.9$ & $  0.59$ & $ 41$ & $  0.51$ & $  74.6$ \\
$  0.90$ & $  0.83$ & $179.7$ & $  0.11$ & $ 16$ & $  0.51$ & $  69.0$ \\
$-18.34$ & $  0.73$ & $169.2$ & $  0.37$ & $134$ & $  0.41$ & $  61.2$ \\
$ 18.07$ & $ -6.62$ & $148.4$ & $  0.04$ & $158$ & $  0.31$ & $  47.1$ \\
$ -5.43$ & $-11.29$ & $141.4$ & $  0.21$ & $ 33$ & $  0.31$ & $  42.8$ \\
$-18.91$ & $-11.12$ & $138.5$ & $  0.41$ & $122$ & $  0.31$ & $  41.0$ \\
$ -9.24$ & $ 21.64$ & $137.6$ & $  0.13$ & $ 45$ & $  0.31$ & $  40.5$ \\
$  7.70$ & $  4.51$ & $136.3$ & $  0.02$ & $151$ & $  0.31$ & $  39.7$ \\
$  8.97$ & $  4.08$ & $133.2$ & $  0.29$ & $ 44$ & $  0.31$ & $  37.9$ \\
$-25.95$ & $  2.73$ & $130.2$ & $  0.02$ & $133$ & $  0.21$ & $  36.2$ \\
$  0.53$ & $ 22.22$ & $122.6$ & $  0.26$ & $ 60$ & $  0.21$ & $  32.1$ \\
$ 17.33$ & $ -2.41$ & $119.8$ & $  0.18$ & $ 60$ & $  0.21$ & $  30.7$ \\
$-22.75$ & $  7.69$ & $119.5$ & $  0.17$ & $ 52$ & $  0.21$ & $  30.5$ \\
$-15.24$ & $ -5.12$ & $111.0$ & $  0.30$ & $ 45$ & $  0.21$ & $  26.4$ \\
$ -0.42$ & $  8.39$ & $105.3$ & $  0.16$ & $ 97$ & $  0.21$ & $  23.7$ \\
$  6.31$ & $  1.85$ & $104.6$ & $  0.36$ & $ 71$ & $  0.21$ & $  23.4$ \\
$  7.85$ & $ -1.99$ & $ 99.2$ & $  0.07$ & $ 19$ & $  0.11$ & $  21.0$ \\
$-13.44$ & $ -9.35$ & $ 97.2$ & $  0.20$ & $ 37$ & $  0.11$ & $  20.2$ \\
$-20.69$ & $  8.54$ & $ 93.9$ & $  0.16$ & $138$ & $  0.11$ & $  18.8$ \\
$  2.30$ & $  9.16$ & $ 81.8$ & $  0.23$ & $127$ & $  0.11$ & $  14.3$ \\
$ -4.95$ & $  7.17$ & $ 86.0$ & $  0.23$ & $ 87$ & $  0.11$ & $  15.8$ \\
\hline
\end{tabular}
\label{tab:lensparam}
\end{table}

\section{Measured properties in the individual lensed images of the $\lowercase{z}=4.88$ arc}
\label{app:table_images}

In \S\ref{sec:arc} we discussed the emission lines properties of the $z=4.88$ arc from the integrated spectrum over galaxy images 1, 2 and 3. In this appendix, we present the constraints that we can measure on the individual galaxy images. In Table \ref{tab:z4.88lines_indiv} we give the redshift and equivalent width measurements of the Ly$\alpha$, \ion{C}{iv}$\lambda$1548{\,\AA} and \ion{C}{iv}$\lambda$1551{\,\AA} emission lines. The individual images (with the exception of galaxy image 1) are too faint to detect the weaker emission lines or the absorption features, while galaxy image 4 is too faint to detect even the \ion{C}{iv} lines at $>3.5\sigma$ and we therefore do not include these lines. We find that the Ly$\alpha$ lines peaks at the same pixel for every image, while the Ly$\alpha$ equivalent width measurements are consistent with each other within the uncertainties. The \ion{C}{iv} EWs are larger for image 2 and 3, which can be explained by the fact that these galaxy images are incomplete and the brightest star-forming region is not included in the measurement.

  \begin{table}
\centering
\caption{Detected spectral features of the $z=4.88$ arc in the individual galaxy images}
\begin{tabular}{lccc}
\hline
\hline

line & $z$ & $\Delta v$~[km$\,\rm s^{-1}$]$\,^a$ & EW$_0$~[\AA] \\
\hline
  \multicolumn{4}{c}{Emission lines}\\ 
\hline
Ly$\alpha^{\rm cont}_{\rm im1}$ & 4.8770$\,^b$ & 68$\pm$37 & 103$\pm$21 \\
Ly$\alpha^{\rm cont}_{\rm im2}$ & 4.8770$\,^b$ & 68$\pm$37 & 141$\pm$27 \\
Ly$\alpha^{\rm cont}_{\rm im3}$ & 4.8770$\,^b$ & 68$\pm$37 & 216$\pm$167 \\
Ly$\alpha^{\rm cont}_{\rm im4}$ & 4.8770$\,^b$ & 68$\pm$37 & 126$\pm$76 \\
\ion{C}{iv}$\lambda$1548{\,\AA}$_{\rm im1}$  & 4.8752  & -23$\pm$26 & 3.5$\pm$0.3 \\
\ion{C}{iv}$\lambda$1548{\,\AA}$_{\rm im2}$  & 4.8750  & -35$\pm$27 & 8.8$\pm$0.8 \\
\ion{C}{iv}$\lambda$1548{\,\AA}$_{\rm im3}$ & 4.8749  & -39$\pm$28 & 6.1$\pm$1.4 \\
\ion{C}{iv}$\lambda$1551{\,\AA}$_{\rm im1}$  & 4.8756  & -6$\pm$26  & 2.5$\pm$0.2\\
\ion{C}{iv}$\lambda$1551{\,\AA}$_{\rm im2}$  & 4.8754 & -15$\pm$26  & 5.3$\pm$0.5\\
\ion{C}{iv}$\lambda$1551{\,\AA}$_{\rm im3}$ & 4.8757  & -2$\pm$28  & 3.9$\pm$1.0\\

\hline

\end{tabular}
\flushleft
$^a$ Velocity offset with respect to the the systemic redshift $z_{[\ion{O}{ii}]}=4.8757\pm0.0005$. Uncertainties combine the uncertainty on the line redshift with the uncertainty on the [\ion{O}{ii}] redshift. 
$^b$ Using the peak of the Ly$\alpha$ line. 
\label{tab:z4.88lines_indiv}
\end{table}

\begin{figure}
\center
\includegraphics[scale=1.1,trim=100mm 195mm 50mm 35mm]{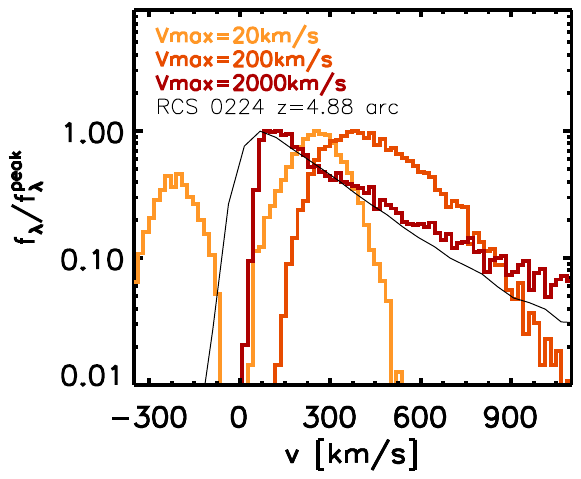}
\caption{The Ly$\alpha$ line profiles from \citet[see their Fig. 7]{Verhamme2006} assuming a velocity gradient $v\propto r$ in the outflowing gas, with three different maximum velocities (20, 200 and 2000 $\rm km\,s^{-1}$), assuming a fixed column density of $2 \times 10^{20}\,\rm cm^{-2}$ and gas temperature of $T = 20000$ K. The black line indicates the Ly$\alpha$ profile of the $z=4.88$ arc presented in this work. The single Ly$\alpha$ peak, emerging close to the systemic velocity and the exponential slope to higher velocities is qualitatively well described by the \citet{Verhamme2006} high velocity ($v_{\rm max}=2000\,\rm km\,s^{-1}$) model. 
}
\label{fig:RCS0224_verhamme}
\end{figure}

\section{Ly$\alpha$ line profiles}
\label{app:lyaprof}
In \S\ref{sec:spec_lya} we described the spectral line shape of the Ly$\alpha$ line observed in the $z=4.88$ arc. The features of this line are well described by the model \citet{Verhamme2006} in which a smooth velocity gradient is present in the outflow of the galaxy. 
In this appendix we present a comparison of the \citet{Verhamme2006} models for velocity gradients with different maximum outflow velocities (20, 200 and $2000\rm\, km\,s^{-1}$, see their Fig. 7) with the Ly$\alpha$ emission from the $z=4.88$ arc in Figure \ref{fig:RCS0224_verhamme}. 
We show spectra as a function of the velocity shift, converting from Doppler units assuming a gas temperature $T = 20000$ K and column density of $2 \times 10^{20}\,\rm cm^{-2}$ as assumed in \citet{Verhamme2006} and we show the spectrum of the  $z=4.88$ arc with respect to $z_{[\ion{O}{ii}]}=4.8757$. 
The main elements of the Ly$\alpha$ emission, such as the single peak emerging close to the systemic velocity and the exponential tail to higher velocities, are present in the model with a high ($2000\,\rm km\,s^{-1}$) maximum velocity and the strongest velocity gradient.

\section{Ly$\alpha$ line emitter candidates}
\label{app:lyaem}

In \S\ref{sec:blind_search} we described our method and testing of a blind line-search for Ly$\alpha$ emitters at $z=4.84-4.92$, $z=5.64-5.80$ and $z=6.54-6.65$. Here we present the $HST$ and MUSE thumbnails of the individual sources and the one-dimensional spectrum from which the sources are identified in figures \ref{fig:RCS0224_lya4.8_1}, \ref{fig:RCS0224_lya5.7_1} and \ref{fig:RCS0224_lya6.3_1}. We also list the sources in table \ref{tab:lya}. We detect asymmetric line profiles in all five sources at $z=4.84-4.92$. However, measuring the continuum flux from the $HST$ imaging using a fixed 0.5$\arcsec$-diameter aperture centered on the Ly$\alpha$ detection, we find only one source (RCS0224\_LAEz4p8773) detected at $>5\sigma$ in the $I_{814}$, $J_{125}$ and $H_{160}$  bands. A slightly weaker signal ($4.6\sigma$) is detected for RCS0224\_LAEz4p8784 in the WFPC2 $V_{606}$ band, while no significant detection ($<2\sigma$) is measured from the redder $HST$ bands of the same source. If real, this flux belongs either to a foreground galaxy or else it would indicate that we have misidentified  RCS0224\_LAEz4p8784 as Ly$\alpha$ emission. However, the asymmetry of the line and the lack of secondary components in the spectrum favour the former interpretation. Furthermore, we find $\sim3\sigma$ detections in the $J_{125}$ band for two sources at $z\sim5.7$, RCS0224\_LAEz5p7405 and RCS0224\_LAEz5p7360. For the 9 Ly$\alpha$ candidates at $z=5.64-5.80$ and $z=6.54-6.65$ we do not have enough S/N to detect the asymmetric profiles, nor do we detect the sources in the $HST$ imaging at $>3.5\sigma$ .

  \begin{table}
\tabcolsep=0.12cm
\centering
\caption{Ly$\alpha$ emitter candidates}
\begin{tabular}{lccc}
\hline
\hline
\newline
ID (short ID$^a$) & RA & Dec & $z_{\rm Ly\alpha}$ \\
\hline
RCS0224\_LAEz4p8556 (l1) & 02:24:36.38 & $-$00:02:17.2 & 4.856 \\
RCS0224\_LAEz4p8784$^b$ (l2) & 02:24:35.93 & $-$00:02:51.2 & 4.878 \\
RCS0224\_LAEz4p8559$^c$ (l3) & 02:24:35.83 & $-$00:02:47.8 & 4.856 \\
RCS0224\_LAEz4p8773 (l4) & 02:24:34.06 & $-$00:02:10.0 & 4.877 \\
RCS0224\_LAEz4p8772 (l5) & 02:24:32.61 & $-$00:02:10.8 & 4.877 \\
\hline
RCS0224\_LAEz5p7093 (l6) & 02:24:36.39 & $-$00:02:49.6 & 5.709 \\
RCS0224\_LAEz5p7412 (l7) & 02:24:35.07 & $-$00:02:18.4 & 5.741 \\
RCS0224\_LAEz5p7845 (l8) & 02:24:35.07 & $-$00:02:18.2 & 5.785 \\
RCS0224\_LAEz5p7362 (l9) & 02:24:33.98 & $-$00:02:06.0 & 5.736 \\
RCS0224\_LAEz5p7405 (l10) & 02:24:33.81 & $-$00:02:14.0 & 5.741 \\
RCS0224\_LAEz5p7360 (l11) & 02:24:33.42 & $-$00:02:16.6 & 5.736 \\
RCS0224\_LAEz5p7169 (l12) & 02:24:33.19 & $-$00:02:56.0 & 5.717 \\
RCS0224\_LAEz5p7352 (l13) & 02:24:32.78 & $-$00:02:19.6 & 5.735 \\
\hline
RCS0224\_LAEz6p6354 (l14) & 02:24:35.22 & $-$00:02:56.8 & 6.635 \\
\hline
\end{tabular}
\flushleft
$^a$ The ID corresponding to the source positions indicated in Figure \ref{fig:RCS0224_color}. 
$^b$ We measure a potential $4.6\sigma$ detection in the $HST$/WFPC2 $V_{606}$ band imaging for this source. 
If associated with the line emission, this detection would indicate a low-redshift solution for the emission line. However, since the line is clearly asymmetric and no secondary components are detected in the spectrum we assume that this flux is associated with a foreground object, or else, that the detected flux arises due to non-Gaussian noise at the edge of the detector.   
$^c$ We measure a $>5\sigma$ detections in the $I_{814}$, $J_{125}$ and $H_{160}$ bands for this source, consistent with this source being an Ly$\alpha$ emitter at $z=4.856$.  
\label{tab:lya}
\end{table}

\begin{figure*}
\includegraphics[scale=1.05,trim=0mm 140mm 60mm 5mm]{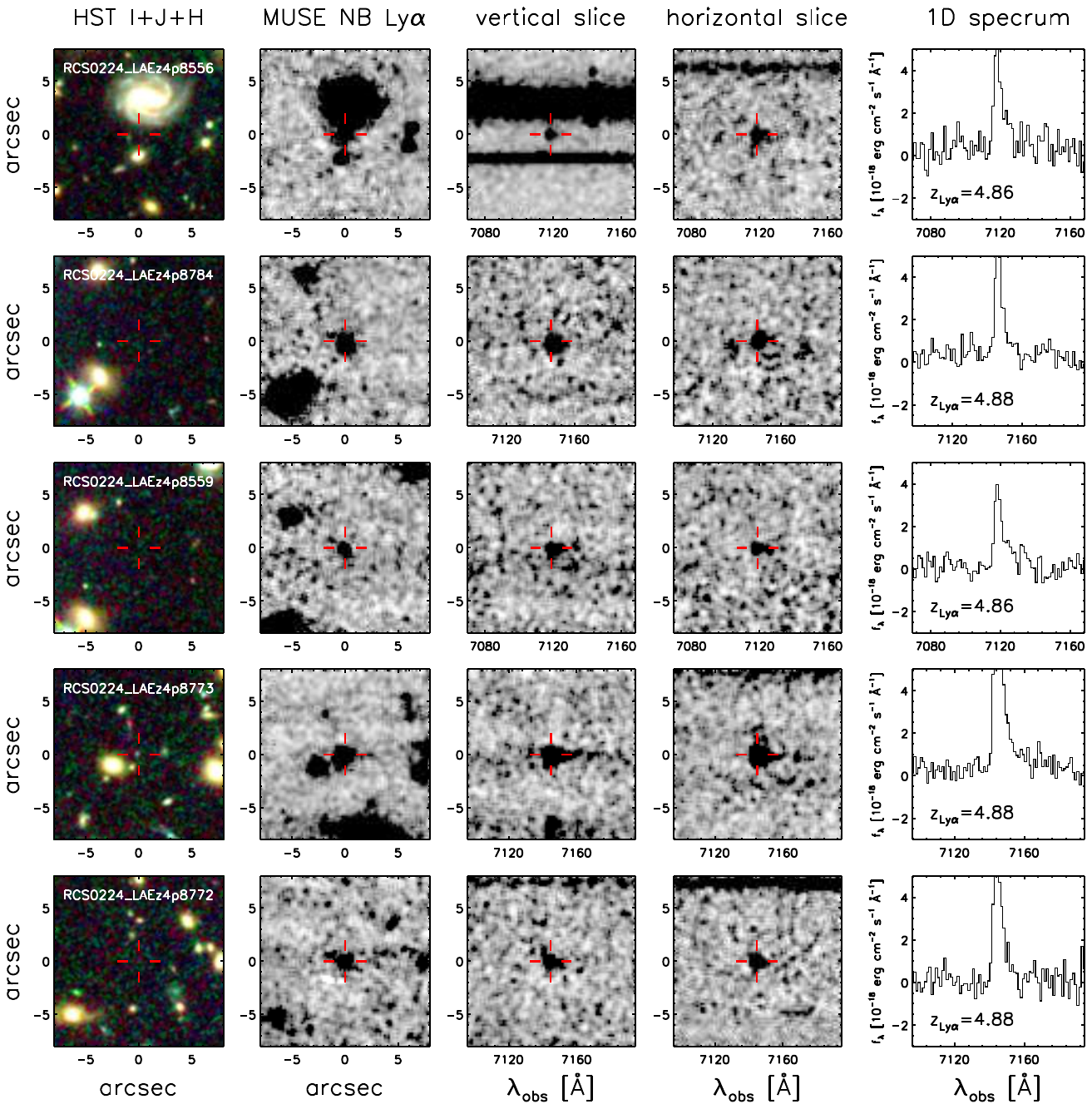}
\caption{ Ly$\alpha$ line emitter candidates in the redshift range $z=4.84-4.92$. From left to right the panels show the combined $HST$ WFPC $V_{606}+I_{814}$ bands (smoothed and rebinned to the MUSE resolution), a narrowband over the FWHM of the identified line in the MUSE cube, a vertical slice from the MUSE cube centered on the line, a horizontal slice of the cube centered on the line and the extracted one-dimensional spectrum from which the line is identified.    }
\label{fig:RCS0224_lya4.8_1}
\end{figure*}

\begin{figure*}
\includegraphics[scale=1.05,trim=0mm 65mm 60mm 5mm]{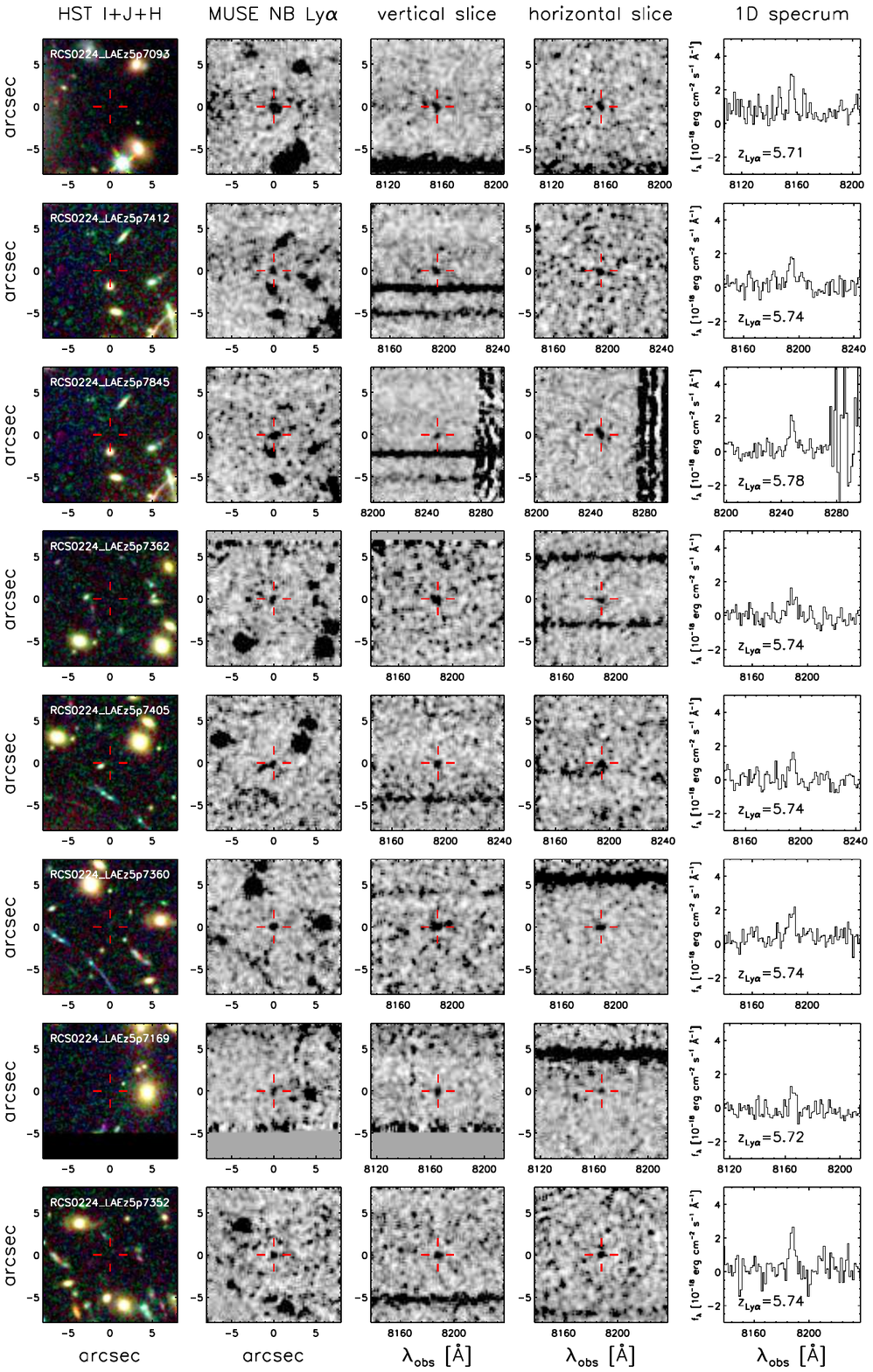}
\caption{ Ly$\alpha$ line emitter candidates in the redshift range $z=5.64-5.80$. Panels are as described in Figure  \ref{fig:RCS0224_lya4.8_1}.    
}
\label{fig:RCS0224_lya5.7_1}
\end{figure*}

\begin{figure*}
\includegraphics[scale=1.05,trim=0mm 240mm 60mm 5mm]{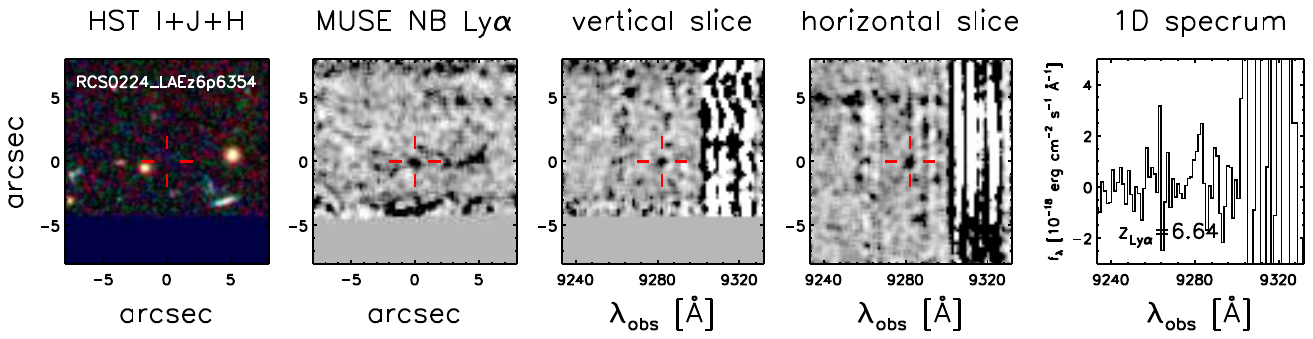}
\caption{ Ly$\alpha$ line emitter candidate in the redshift range $z=6.54-6.65$. Panels are as described in Figure \ref{fig:RCS0224_lya4.8_1}. }
\label{fig:RCS0224_lya6.3_1}
\end{figure*}

\end{document}